\documentclass[twocolumn,prd,floatfix,preprintnumbers,a4paper,nofootinbib,superscriptaddress,groupedaddress]{revtex4-1}
\usepackage[utf8]{inputenc}
\usepackage{amsmath,amssymb}
\usepackage{amsfonts}
\usepackage{bm}
\usepackage{graphics}
\usepackage{float}
\usepackage{amsmath}
\usepackage{amssymb}
\usepackage[normalem]{ulem} %for \sout
\usepackage[mathscr]{euscript}
\usepackage{acronym}
\usepackage{float}
\usepackage[caption=false]{subfig}
\usepackage{lipsum}
\usepackage{mathrsfs}
\usepackage{mathtools}
\usepackage{multirow}
\usepackage{graphicx}
\usepackage{mathtools}
\usepackage{multirow}
\usepackage{graphicx}
\usepackage[usenames,dvipsnames,table,xcdraw]{xcolor}
\usepackage[colorlinks, pdfborder={0 0 0}]{hyperref} 
\usepackage[nameinlink,capitalise]{cleveref}
\usepackage{xspace}
\usepackage{xfp}

\graphicspath{{fig}}
\newcommand{\todo}[1]{\textcolor{black}{#1}}

\newcommand{\xj}[1]{\todo{#1}}

\graphicspath{{fig}}

\newcommand{\macro}[1]{#1}   
\newcommand{\Nrcasesalln}{\macro{620}}		
\newcommand{\Nrcasesn}{\macro{610}}
\newcommand{\Nrcasesnn}{\macro{18}}
\newcommand{\Nrresoln}{\macro{343}}
\newcommand{\Nrresol}{\macro{\Nrresoln\xspace}}
\newcommand{\Outliers}{\macro{$\fpeval{\Nrcasesalln-\Nrcasesn}$\xspace}}
\newcommand{\Nrcasesall}{\macro{\Nrcasesalln\xspace}}
\newcommand{\Nrcases}{\macro{\Nrcasesn\xspace}}
\newcommand{\Nrcasesneg}{\macro{\Nrcasesnn\xspace}}

\newcommand{\epsone}{\macro{10^{-1}}}	
\newcommand{\epsthree}{\macro{1.3 \times 10^{-2}}}	
\newcommand{\epsseven}{\macro{2.2\times 10^{-3}}}

\newcommand{\epshigh}{\macro{2\times 10^{-3}}}	
\newcommand{\epserrl}{\macro{4.6\times 10^{-5}}}	
\newcommand{\epserrr}{\macro{8.1\times 10^{-4}}}	
\newcommand{\epshighmin}{\macro{6\times 10^{-4}}}	

\newcommand{\spinmin}[1]{a_{f,#1}^\mathrm{start}}

\newcommand{\nmax}{N_{\mathrm{max}}}
\newcommand{\sxs}[1]{SXS:BBH:#1}

\begin{document}

\title{High-overtone fits to numerical relativity ringdowns: Beyond the dismissed $n=8$ special tone.}
% The missed special tone and beyond.
%%%
%\author{Xisco Jim\'enez Forteza$^{1}$,
%Pierre Mourier$^{2}$}
\author{Xisco Jim\'enez Forteza$^{1,2}$,
Pierre Mourier$^{1,2}$}

\affiliation{$^1$ Max Planck Institute for Gravitational Physics (Albert Einstein Institute), Callinstra{\ss}e 38, 30167 Hannover, Germany}
\affiliation{$^2$  Leibniz Universit\"at Hannover, 30167 Hannover, Germany}

\begin{abstract}
In general relativity, the remnant object originating from an uncharged black hole merger is a Kerr black hole.
%This state is reached by emitting
This final state is reached through the emission of a late train of radiation known as the black hole ringdown. In linear perturbation theory around the final state, the ringdown morphology is described by a countably infinite set of damped sinusoids --- the quasinormal modes --- whose complex frequencies are solely determined by the final black hole's mass and spin. Recent results advocate that ringdown waveforms from numerical relativity can be fully described from the peak of the strain onwards if quasinormal mode models with $\nmax=7$ overtones (beyond the fundamental mode) are used. In this work we extend this analysis to models with $\nmax \geq 7$ up to $\nmax=16$ overtones by exploring the parameter bias on the final mass and spin obtained by fitting the nonprecessing binary black hole simulations from the SXS catalog. To this aim, we have computed the spin weight $-2$ Kerr quasinormal mode frequencies and angular separation constants for the $(l=m=2, n=8,9)$ co- and counter-rotating overtones, which all approach a Schwarzschild algebraically special mode at low spins. We provide tables of the values obtained for these modes, which are in agreement with previous results. From the systematic variable-$\nmax$ analysis, we find that $\nmax\sim 6$ overtones are on average sufficient to model the ringdown from the peak of the strain, although about $21\%$ of the cases studied require at least $\nmax\sim 12$ overtones to reach a comparable accuracy on the final state parameters. Considering the waveforms from an earlier or later point in time, we find that a very similar maximum accuracy can be reached in each case, occurring at a different number of overtones $\nmax$. We also provide new error estimates for the SXS waveforms based on the extrapolation and the resolution uncertainties of the gravitational wave strain, which dominate over the errors obtained from the quasilocal measures of the final mass and spin. Finally, we observe substantial instabilities on the best-fit amplitudes of the tones beyond the fundamental mode and the first overtone, that, nevertheless, do not impact significantly the mass and spin estimates.
%Those values are publicly available.
\end{abstract}
%%%%%%%%%%%%%%%%%%%%%%%%%%%%%%%%%%%%%%%
\maketitle
%%%%%%%%%%%%%%%%%%%%%%%%%%%%%%%%%%%%%%%%
%%%%%%%%%%%%%%%%%%%%%%%%%%%%%%%%%%%%%%%%%%
\section{Introduction}
The number of gravitational wave (GW) observations is increasing along with the upgrades of GW interferometers. Up to date, the LIGO-Virgo collaboration has reported a total of 48  binary black hole merger candidates~\cite{gwtc1,gwtc2}. Those observations are providing unprecedented constraints on general relativity in its strong-field regime, with the merger-ringdown phase in particular providing a promising channel for such studies.

A binary black hole merger is generally decomposed in three different regimes that depict its orbital evolution: inspiral, merger and ringdown. The inspiral regime represents the \emph{slow} far-field solution and it is well described by post-Newtonian and effective-one-body theories. At the merger phase, that is, when the two bodies get closer to each other, these analytic solutions break down due to the strong general-relativistic effects and full numerical relativity is needed. The final merger results in an initially perturbed space-time that evolves towards the final Kerr solution by emitting an ultimate tail of radiation better known as ringdown (RD). The strain $h(t,\theta,\phi)$ of the RD waveform is predicted by linear perturbation theory to decompose as a sum of damped sinusoids:
\begin{align}
\label{eq:rdmodel}
    h(t,\theta,\phi) = \sum_{l,m,n} {\mathcal A}_{lmn} e^{- \iota \omega_{lmn} (t-t_r)}\, {}_{-2}\mathcal{Y}_{lm}(\theta,\phi)\, , \, t \geq t_r \, .
\end{align}
Here, $l=2,3,\dots$ and $m=-l,-l+1, \dots, l-1, l$ account for the two angular indices of the spheroidal decomposition, while $n=0,1,2,\dots$ labels the tone; $_{-2}\mathcal{Y}_{lm}(\theta,\phi)$ are the spin-weighted spheroidal harmonics of spin weight $s = -2$, as functions of the polar angle $\theta$ and azimuthal angle $\phi$; ${\mathcal A}_{lmn}=A_{lmn} \, e^{\iota \varphi_{lmn}}$ is the tone complex amplitude; and $t_r$ is some undefined time beyond which linear perturbation theory is expected to accurately describe the RD regime~\cite{giesler2019,london:2014cma,Bhagwat:2019dtm}. In particular, and for non-charged black holes, the $\omega_{lmn}=w_{lmn}-\iota/\tau_{lmn} $ defines a countably infinite set of complex frequencies solely determined by the final black hole's mass $M_f$ and spin $a_f$, where the values of $\omega_{lmn}$  correspond to poles of the Green function to the inhomogeneous Teukolsky equation --- the quasinormal modes (QNMs) of the final black hole~\cite{leaver:1985ax,detweiler:1980gk,kokkotas:1999bd}. Here $\mathrm{Re}[\omega_{lmn}]=w_{lmn}$ and $-\mathrm{Im}[\omega_{lmn}]=1/\tau_{lmn}$ take the role of the oscillation frequency and the damping rate (inverse of the damping time) respectively. As a rule of thumb, if one considers fixed the value of the $(l,m)$ indices, the mass $M_f$, and for moderate spins $a_f$, the values of the damping times $\tau_{lmn}$ decrease as the tone index $n$ increases. This sets the $n=0$ (fundamental) tone as the dominant tone while the $n \geq 1$ tones (overtones) rank down continuously as $n$ increases. Moreover, one finds two branches of solutions for $\omega_{lmn}$ also known as the corotating (dominant \xj{for $m>0$}) and counter-rotating (subdominant \xj{for $m>0$}) modes~\cite{Finch:2021iip,Dhani:2020nik,berti:2009kk,forteza2020,cook2020,Dhani:2021vac} which both contribute\footnote{%
Such contribution involves an additional sum over a binary index labeling the co- and counter-rotating QNM frequencies and amplitudes which have been dropped here to simplify the notation. In the following we shall rather explicitly state, when needed, whether a given QNM is associated to a co- or counter-rotating branch. Note that the counter-rotating modes excited in a binary black hole merger are usually expected to have negligible amplitudes compared to the corotating modes \xj{(for $m > 0$ harmonics)}~\cite{forteza2020,Finch:2021iip}.%
} to Eq.~\eqref{eq:rdmodel}. \xj{Co- and counter-rotating modes are distinguished by the sign of the real part $w_{lmn}$ of their frequencies. While multiple conventions exist, in this work we follow the convention of, \emph{e.g.}, \cite{berti:2009kk,Cook:2014cta}, and we denote as corotating the family of modes with positive $w_{lmn}$ regardless of the sign of $m$ or of the sign attributed to the black hole's spin $a_f$.} 

The black hole no-hair and uniqueness theorems in general relativity imply that the final state of an uncharged black hole merger, and the associated QNM spectrum, are uniquely determined by the values of the final mass and spin. This has led to two main avenues to test such theorems. The first one consists on performing an inspiral--merger--ringdown (IMR) consistency test, which relies on independently estimating the final black hole mass and spin from both the inspiral--merger and the ringdown phases~\cite{LIGOScientific:2020tif}. The second approach is to perform black hole spectroscopy, which typically aims at independently estimating the parameters of the fundamental tone of the dominant angular mode, $(l=2,m=2,n=0)$, plus at least another mode, either i) the first corresponding overtone, $(l=2,m=2,n=1)$, or ii) another angular fundamental mode, either the $(l=m=3,n=0)$ or the $(l=2$, $m=1$, $n=0)$ mode (in order of importance). So far and for unequal-mass-ratio binaries, the higher angular mode remains the most promising approach to test the implications of the black hole no-hair theorem~\cite{forteza2020}. Successful independent evidence of the $(l=m=2,n=0)$ and the $(l=m=3,n=0)$ modes in the ringdown phase of a GW event (in this case GW190521) has been recently provided in~\cite{Capano:2021etf}. On the other hand, channel i) becomes a promising possibility when dealing with near equal-mass-ratio nonspinning binaries. For such events, the higher harmonic modes are only weakly excited, while the overtones would still represent a valid channel in the ringdown regime. A first attempt to observe overtones in GW observational data has been performed in~\cite{isi2019,Isi:2021iql} on GW150914. However, the full spectroscopic analysis performed by~\cite{Capano:2021etf} on GW190521 could not find evidence of tones other than the fundamental ones.

Current studies of the $(22n)$ ringdown modes rely on fits to numerical relativity (NR)  waveforms~\cite{Bhagwat:2019dtm,forteza2020,giesler2019,london:2018nxs}, which are shown to be consistent with current GW observations. In particular, using NR waveforms has the following advantages: i) the underlying theory is well-known; ii) the mass and the spin of the final black hole (BH) are accurately estimated, hence accurately determining the QNM spectrum; and iii) numerical errors in the simulated waveforms are typically smaller than current GW detectors noise. In such studies, considering the $(l=m=2)$ spherical harmonic of the strain\footnote{It is worth mentioning here that in NR codes, the strain $h(t,\theta,\phi)$ is decomposed in terms of the spin-weighted spherical harmonics basis instead of the spheroidal harmonics $\mathcal{Y}_{lm}$ used to define QNMs, since it is a better adapted basis to the inspiral-merger regime. This adds mode-mixing artifacts between both bases principally at modes other than the $(22)$ mode~\cite{berti:2014fga,cook2020}.}, $h_{22}(t)$, from a given time $t_0$ onwards, one fits for the successive complex amplitudes $\mathcal{A}_{22n}$ of the $(22n)$ QNM tones for a running index $n \in \{ 0, \dots, \nmax \}$, with various choices for the total number $\nmax$ of overtones to be included in the model. The $\nmax=7$ model has been shown to provide the best estimates of the true final parameters (mass and spin)~\cite{giesler2019,Finch:2021iip}, although no models beyond $\nmax = 7$ have been studied up to date. In this work we extend this analysis to $\nmax \geq 7$. 
In particular, we have not found any publicly available catalog of Kerr QNM data that provides a correct description for the $n=8$ tone nor for the neighboring corotating tone that we label as $n=9$ here\footnote{%
The catalogs~\cite{berti:2005ys,berti:2009kk,berti-webpage,vitor-webpage} and \cite{Cook:2014cta,cook-QNM} provide the QNM solutions up $n=7$. On the other hand, the solutions provided by the \texttt{qnm} Python package~\cite{Stein:2019mop} up to much larger $n$ values are incorrect at ${n=8}$ (both for the co- and counter-rotating modes) due to the erroneous estimate of the Schwarzschild limit, and are missing the neighboring corotating branch that we label here as $n=9$. Solutions for these $n=8,9$ modes have been previously obtained by
~\cite{Onozawa_1997} ($n=8$ modes only) and~ \cite{Berti:2003,Cook:2014cta,Cook:2016A,Cook:2016B} (with a different labeling of what we here call the $n=8,9$ corotating modes as two corotating $n=8$ branches), where they are shown as $\mathrm{Re}(\omega_{lmn})$ --- $\mathrm{Im}(\omega_{lmn})$ frequency plots but the data obtained for $n\geq 7$ modes were not made publicly available.
}. We discuss these tones further in Sec.~\ref{sec:QNM} below and show our results for their frequencies in Sec.~\ref{sub:qnmresults}, comparing to the results from~\cite{Onozawa_1997,Berti:2003,Cook:2014cta,Cook:2016A}. In Sec.~\ref{sec:nr_data} we revisit the definitions of the mass and spin in NR simulations and we provide two methods to compute their uncertainties. Finally, in Sec.~\ref{sec:fits}, we show the results on the mass and the spin obtained from fitting models with up to $\nmax=12$ or $16$ overtones to NR waveforms from the SXS and RIT catalogs~\cite{sxscatalog,ritcatalog} while we further discuss the model instabilities.
\section{The ringdown QNM spectrum}
\label{sec:QNM}

\subsection{The ringdown wave equation}
The Teukolsky master equation~\cite{teukolsky:1974yv} describes the propagation of linear perturbations of fields of general spin weight $s$ in a Kerr background~\cite{chandrasekhar:1985kt,berti:2009kk,kokkotas:1999bd}. The angular ($\mathcal{Y}_{lm}$) and radial ($R_{lm}$) sector of this equation read, respectively,
\begin{align}
&\frac{\partial }{\partial u} \left[\left(1-u^2\right)\frac{\partial}{\partial u} \,  {}_{s}\mathcal{Y}_{lm}\right] + \left[(a_f\, \omega \,u)^2 \phantom{\frac{\left(m+ su\right)^2}{1-u^2}} \right. \nonumber \\
{} -& 2 \, a_f\,\omega\,s\, u +s + \left. {\mathscr{A}}-\frac{\left(m+ su\right)^2}{1-u^2}\right] {}_{s}\mathcal{Y}_{lm}=0 \, ; \nonumber \\
&\Delta \, \partial^2_r R_{lm}+(s+1)(2r-2M_f)\partial_r R_{lm}+ V R_{lm}=0\,,
    \label{eq:teuks}
\end{align}
where ${s=\pm 2}$ for gravitational perturbations. Here $u=\cos \theta$; $\Delta =(r-r_-)(r-r_+)$, $r$ is the radial coordinate while $r_{+,-}$ stand for the coordinate radii of the outer and inner BH horizons respectively; $l,m$ are the usual angular indices; $a_f$ is the black hole's spin and $M_f$ the black mass; $V=V(r,M_f,a_f,\omega,\mathscr{A},s,m)$ is the potential term for a Kerr BH (see Eq.(26) of~\cite{berti:2009kk}); $\omega$ is the complex frequency of the perturbation; and $\mathscr{A}$ is the corresponding so-called angular separation constant.%\footnote{It is worth mentioning here that in NR codes Eq.~\eqref{eq:rdmodel} is decomposed in terms of the spin-weighted spherical harmonics basis instead of the spheroidal harmonics $\mathcal{Y}_{lm}$ used to define QNMs, since it is a better adapted basis to the inspiral-merger regime. This adds mode-mixing artifacts principally at modes other than the $(22)$ mode~\cite{berti:2014fga,cook2020}.}

For each value of the final spin $a_f$, each spin weight $s$ and each angular mode $(l,m)$, the $(l,m,n)$-- quasinormal modes are obtained by imposing outgoing boundary conditions at spatial infinity and ingoing boundary conditions at the black hole horizon. The QNMs and associated angular separation constants form the only discrete set of (complex) values $\left\lbrace \omega,\mathscr{A} \right\rbrace$ that are compatible with these
boundary conditions. Eq.~\eqref{eq:teuks} can be solved and its associated QNMs values obtained following the algorithm proposed in~\cite{leaver:1985ax}.  The frequency and separation constant solutions are then labeled by the integers $l$, $m$ and $n$: $\omega \equiv \omega_{lmn}$,  $\mathscr{A} \equiv \mathscr{A}_{lmn}$, where $n=0,1,2,\ldots$ is the overtone index. The dependence on the spin weight $s$ usually remains implicit; we only consider gravitational perturbations here and we set $s=-2$ throughout this work. See
\cite{leaver:1985ax} for a method for numerically calculating the
QNM spectrum, \cite{berti-webpage,vitor-webpage,berti:2005ys,berti:2009kk} and \cite{cook-QNM,Cook:2014cta} for a compilation of the values in
different situations and up to $n=7$ for the Kerr scenario, and \cite{Stein:2019mop}
 for a Python package, \texttt{qnm}, to evaluate the QNM spectrum of Kerr black holes for a variety of $(l,m,n)$ modes and spin weights $s$.
%The metric perturbations propagating in a Kerr space-time are described by the solution of the Teukolsky equation,
%\begin{equation}
%\label{eq:teukolsky}
%  \frac{d^2\psi}{dr_\star^2} + \nu^2\psi = V_{\pm}\psi\,,
%\end{equation}
%where  $r_\star = r + 2M_f \log(r/2M_f - 1)$, $r$ is the Schwarzschild radius and $M_f$ is the black hole final mass in geometric units. The $V_{\pm}$ describes the  gravitational potential for axial and polar perturbations respectively and it depends on the 
%The RD strain takes the following form,
%\begin{align}
%\label{eq:rdmodel}
%    h(t) = \Sigma_{lmn} {\cal C}_{lmn} e^{- \iota \omega_{lmn} t} e^{- (t-t_0)/ \tau_{lmn}}\, _{s}\mathcal{Y}_{lm}\,,
%\end{align}
%where ${\cal C}_{lmn}=A_{lmn}e^{\iota \phi_{lmn}}$ and  $\tau_{lmn}$ are the quasinormal-mode~(QNM) frequencies and damping time respectively. The $(l,m)$ indices describe the angular decomposition of the modes, $_s\mathcal{Y}_{lm}$ are the spin-weighted $s=-2$ spheroidal harmonics, $n$ accounts for the $n$-tone excitations of a given $(l,m)$ mode, with $n=0$ being the fundamental ``tone''. 

\subsection{Computing the ($22n$) quasinormal modes}
\label{sub:qnmcode}

Most of the QNM frequency values used for this work (which all correspond to the $(l=2,m=2)$ harmonic) were computed using the dedicated \texttt{qnm} Python package \cite{Stein:2019mop}. However, the method used (Leaver's method~\cite{leaver:1985ax}) is known to fail for the $(l=m=2, n=8)$ mode in the Schwarzschild limit as it becomes an algebraically special mode~\cite{chandrasekhar:1975zza,Berti:2003}. Hence, this tone was flagged as unreliable in this code since the spectrum computation relies on the Schwarzschild limit.  Indeed, the results from \texttt{qnm} at any spin for this mode appear to be inconsistent with the neighboring modes, and are in disagreement with~\cite{Onozawa_1997,Berti:2003,Cook:2014cta}, for both the co- and counter-rotating branches. Hence, we rather computed the QNM frequencies $\omega_{lmn}$ --- along with the angular separation constants $\mathscr{A}_{lmn}$ --- for this mode from a modified version of the publicly available Mathematica code for Kerr QNMs from \cite{berti:2005ys,berti:2009kk} (available online at \cite{berti-webpage,vitor-webpage}).

This code is also based on Leaver's continued fraction method \cite{leaver:1985ax} --- but it may be used to directly compute the QNMs for any given spin, without relying on the Schwarzschild limit. Following this method, estimates of $\omega_{lmn}$ and $\mathscr{A}_{lmn}$ are found successively as roots of infinite generalized continued fractions, which are approximated with a finite numbers of fraction steps $n_{\mathrm{frac}}$. The coefficients involved in the fraction used for each of both variables depend on the estimate of the other variable, thus the alternated estimation of $\omega_{lmn}$ and $\mathscr{A}_{lmn}$ is iterated until convergence is reached. We have modified the continued fraction computation to include Leaver's inversions (Eq.~(14) in \cite{leaver:1985ax}) allowing for a more stable recovery of any given overtone ($n \geq 1$), and replaced the use of Mathematica's time- and memory-consuming root-finding algorithm by a direct implementation of the secant method. We ensure the convergence both in terms of $n_{\mathrm{frac}}$ and of the $(\omega_{lmn}, \mathscr{A}_{lmn})$ loop by increasing $n_{\mathrm{frac}}$ by a constant factor $c_{\mathrm{frac}} > 1$ at each iteration, until a convergence criterion is met. This criterion amounts to requiring that the (absolute) variations of the estimates of both variables over three consecutive iterations do not exceed a certain threshold (which we set at $3 \cdot 10^{-11}$). This progressive increase of $n_{\mathrm{frac}}$ was necessary for the computation of the $n=8$ modes (as well as the neighboring corotating branch which we label here as $n=9$), as these modes typically require rather large values of $n_{\mathrm{frac}}$ (further increasing as the spin gets closer to zero or one) to reach such an accuracy; we set $c_{\mathrm{frac}} = 1.2$ for this computation. We have checked for consistency that this algorithm provides the same $s=-2$ solutions as those for available $(l=2,m=\pm 2, n \neq 8)$ tones from~\cite{berti:2005ys,berti:2009kk,berti-webpage,vitor-webpage,Stein:2019mop}.

 \xj{We make this modified Mathematica code available here:~\cite{my-codeberg, my-github}, along with its translation into Python and into Fortran. We also made use of this (much faster) Fortran equivalent, to obtain the frequency solutions for these modes at a few spin values in the regimes where convergence to the solution is particularly difficult to achieve.}

The algorithm does require an initial guess for $\omega_{lmn}$\footnote{%
 More precisely, since we modified the code to use the secant method rather than Newton's method to find the roots of the continued fractions, two initial guesses on $\omega_{lmn}$ are required instead of one. These may simply be chosen as two close yet distinct estimates such as bounds on the expected solution or simply perturbations around a given estimate. This is also required for initial guesses on $\mathscr{A}_{lmn}$. For this variable we used systematic small deviations above and below the single guess value that was originally used to initialize Newton's method --- that is, at each iteration beyond the first, the $\mathscr{A}_{lmn}$ result from the previous iteration, and at the first iteration, the Schwarzschild-limit solution $\mathscr{A}_{lmn} = l(l+1) - s(s+1)$.
}, which simply needs to  lie closer to the desired mode than to any other tone of the same $(l,m)$ harmonic --- in practice for the modes discussed here, a $\sim 5 \%$ accurate initial estimate is typically sufficient. This allowed us to recover the $n=8$ co- and counter-rotating QNMs over a wide spin range (see below) by simply using initial guesses based on a few points of the $n=8$ curves in Fig.~4 of \cite{Onozawa_1997} and interpolation and extrapolation between and beyond them. Using slightly lower imaginary values for the initial guesses on the frequency with respect to the above corotating solution, we also recovered the additional corotating mode that also nears the imaginary axis at low spin identified by~\cite{Berti:2003,Cook:2014cta} and missing in~\cite{Onozawa_1997} and in the \texttt{qnm} package.
\vspace{-1ex}

\subsubsection*{Remarks on the tone labeling convention}
\vspace{-1.5ex}
We label in this work this additional corotating branch as the $n=9$ mode and the subsequent ones (with even smaller values of $\mathrm{Im}[\omega_{lmn}]$) as $n=10, 11, \dots$
The counter-rotating mode associated with each of the $n \neq 8,9$ corotating tones, connecting to the same Schwarzschild limit up to a $\mathrm{Re}(\omega) \mapsto - \mathrm{Re}(\omega)$ symmetry, is attributed the same $n$ index. This overall leads to an unusual convention for $n \geq 8$ and can be somewhat confusing: our $(l=2,m=2,n=n_0)$ co- and counter-rotating QNMs for each $n_0 \geq 10$ are equivalent to the $(l=2,m=2,n=n_0-1)$ solutions for the \texttt{qnm} package~\cite{Stein:2019mop} or from~\cite{Cook:2016A} for instance, and in the Schwarzschild limit, they match the Schwarzschild QNM that is usually attributed the overtone index $n_0-1$ in the literature.
This offset is due to the presence of a single Schwarzschild ($l=2,m=2$) QNM, traditionally labeled $n=8$, at $\omega = -2 \, \iota$ --- which coincides with an algebraically special mode
~\cite{MaassenvandenBrink:2000,Cook:2014cta} --- while two distinct Kerr QNMs are found near this value at low spins~\cite{Berti:2003,Cook:2014cta}. This is reconciled in~\cite{Berti:2003} by simply considering these two Kerr solutions as a double $n=8$ branch. The same choice is made in~\cite{Cook:2014cta,Cook:2016A,Cook:2016B} where, specifically, these Kerr modes that we here refer to as $n=8$ and $n=9$ overtones are labeled as the $n = 8_0$ and $n = 8_1$ modes, respectively.
On the other hand, the convention that we adopt here allows for a consistent sorting of the found Kerr QNMs by decreasing imaginary frequency (or decreasing damping time) for a given spin, and for the preservation of a roughly equal spacing between $\mathrm{Im}[\omega_{l,m,n}]$ and $\mathrm{Im}[\omega_{l,m,n+1}]$ for corotating modes for all values of $n$, over most of the spin range.

\bigskip{}

This latter property could in fact be used to get reasonable initial guesses for the frequency values of both the $n=8$ and $n=9$ corotating modes over most of the spin range ($a_f \gtrsim 0.1$), without prior knowledge about these values, by extrapolating the neighboring tones' frequencies as a function of $n$ for each $a_f$. In particular, we did generate initial guesses in this way to systematically compute the $n=9$ mode frequencies for spins $a_f \geq 0.1$. Initial guesses for lower spin values for this mode were obtained by successive extrapolations of the results previously obtained at larger spins.

\subsection{The $\omega_{22n}$ curves}
\label{sub:qnmresults}
\begin{figure}[h]
\includegraphics[width=0.98\columnwidth]{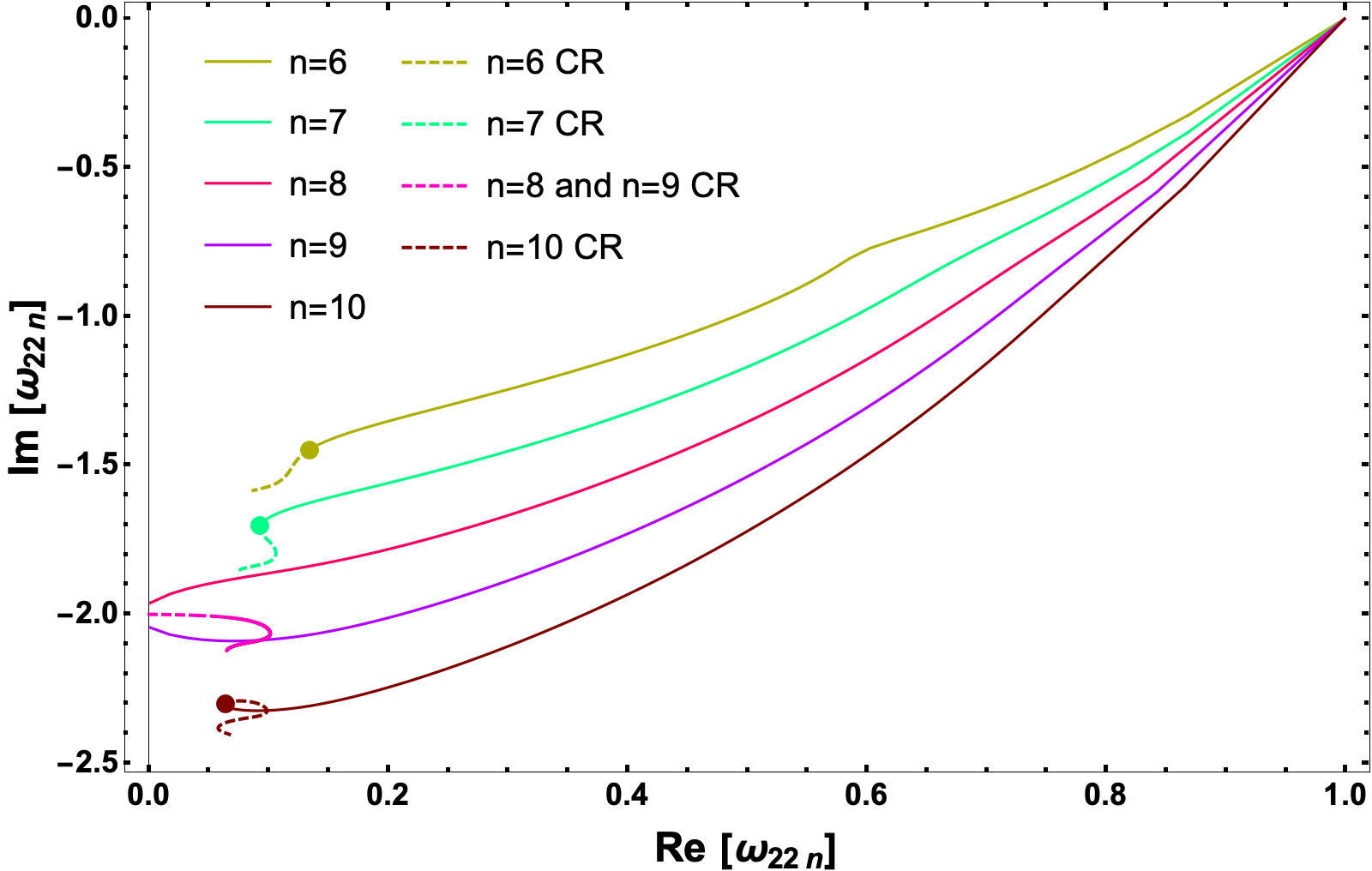}
    \caption{Co- and counter-rotating (CR) QNM frequencies $\omega_{lmn}$ on the complex plane for a final black hole mass set to $M_f = 1$, for $n\in \left\{6,\dots,10\right\}$ and for spin $a_f$ spanning $[0, 1]$ or a wide subset of this range (as discussed in the main text, Sec.~\ref{sub:qnmresults}). The solid lines account for the corotating solutions while the dashed ones correspond to the counter-rotating tones. The latter are represented under the $\omega \mapsto - \omega^*$ transformation, or equivalently, as per Eqs.~\eqref{eq:flipm}--\eqref{eq:sym}, as corotating modes with either $m=2 \mapsto m=-2$, or, $a_f \mapsto -a_f$. For $n \neq 8,9$, the Schwarzschild limit is recovered at the joining point between the dashed and solid lines, and is materialized by a dot. At the extremal-Kerr limit for all ($a_f > 0$) corotating modes shown here, $\omega_{lmn} \rightarrow 1$ as expected. The roughly equal spacing in $\mathrm{Im}(\omega_{22n})$ between successive ($a_f > 0$) corotating tones can be noted on these curves for $\mathrm{Re}(\omega_{22n}) \gtrsim 0.2$, corresponding to spins $a_f \gtrsim 0.3$. This holds down to smaller spins $a_f \gtrsim 0.1$ up to small shifts in $\mathrm{Re}(\omega_{22n})$ at fixed $a_f$ between the tones shown here.}
    \label{fig:wlmn}
\end{figure}

In Fig.~\ref{fig:wlmn} we show the range of $\omega_{lmn}$ values for the $(l,m)=(2,2)$ and $n\in \left\{6, \dots, 10 \right\}$ corotating (solid curves) and counter-rotating modes (dashed curves) for a unit-mass ($M_f = 1$) Kerr black hole as its dimensionless spin $a_f$ varies. These curves correspond to the full range of spin values $a_f \in [0, 1]$ for $n\not= 8,9$, and to the ranges over which we could obtain solutions for the $n=8,9$ tones: $a_f \in \left[\spinmin{n}, 1\right]$ for the corotating modes, with $\spinmin{8} \equiv \xj{3.5} \cdot 10^{-3}$ for $n=8$ and $\spinmin{9} \equiv \xj{5.4} \cdot 10^{-3}$ for $n=9$, and $a_f \in \xj{ \left[0, 0.997\right] }$ for the associated counter-rotating mode.
The counter-rotating mode curves shown have actually been flipped around the imaginary axis (using the symmetry transformation $\omega \mapsto -\omega^*$ where ${}^*$ stands for the complex conjugation) for better visualization. This choice can alternatively be interpreted as a representation of the corresponding corotating solutions either for the $(l=2,m=-2)$ mode (as in~\cite{Berti:2003,Cook:2014cta,Cook:2016A,Cook:2016B}) or for negative spins, from the following symmetry relations~\cite{berti:2009kk,Cook:2014cta}:
\begin{align}
\label{eq:flipm}
\omega_{lmn}\left(a_f\right) = -\omega_{l(-m)n,c}^*\left(a_f\right) \, ; \\
\omega_{lmn}\left(a_f<0\right) = -\omega_{lmn,c}^*\left(|a_f|\right) \, ,
\label{eq:sym}
\end{align} 
where $ \omega_{{l(\pm m)n},c}$ stand for the counter-rotating mode frequencies.
The negative-spin interpretation explains that most of these curves continuously connect to the ($a_f \geq 0$) corotating branches.
The Schwarzschild limit for each $n \neq 8,9$ mode then appears at the transition point between dashed and solid curves in the figure and is marked with a dot; while the extremal-Kerr limit $a_f \rightarrow 1$ is found as $\left\lbrace \mathrm{Re}[\omega_{lmn}],\mathrm{Im}[\omega_{lmn}] \right\rbrace \rightarrow \left\lbrace 1,0 \right\rbrace$ for all of the corotating modes shown here (\emph{cf.}~\cite{chandrasekhar:1975zza}).

For the corotating branch, $\mathrm{Re}[\omega_{lmn}]$ increases monotonically with the spin of the final object $a_f$. Similarly, $\mathrm{Im}[\omega_{lmn}]$ increases with $a_f$ over most of the spin range for $n\geq 9$ and over the whole range for $n\leq 8$. Moreover, one can notice that the values of the corotating modes span a broader domain (both in $\mathrm{Re}[\omega_{lmn}]$ and $\mathrm{Im}[\omega_{lmn}]$) with respect to the counter-rotating branch. The same features apply as well for all tones other than those shown, except for $n=5$ which has a different high-spin behavior~\cite{Onozawa_1997}. Regarding the counter-rotating branches, the solutions decrease monotonically with $a_f$ both in real and imaginary part up to $n=6$. For $n\geq 7$ the solutions become degenerate in $\mathrm{Re}[\omega_{lmn}]$, where one can find multiple values of $\mathrm{Im}[\omega_{lmn}]$ given one fixed $\mathrm{Re}[\omega_{lmn}]$.

While the other modes were computed using the \texttt{qnm} Python package, as mentioned above the $n=8$ (co- and counter-rotating) and $n=9$ curves that we present in this figure have been obtained from the adapted version of the Mathematica code\footnote{%
We also used this code to complete the curves for the other corotating modes up to spins close to $1$, as the \texttt{qnm} package results become unreliable ---typically swapping different tones--- at very high spins ($a_f \gtrsim$ 0.995).
} from \cite{berti:2005ys,berti:2009kk,berti-webpage,vitor-webpage} described in Sec.~\ref{sub:qnmcode}.  
Our results for these three modes are in good qualitative agreement with those shown in~\cite{Onozawa_1997} (limited to the $n=8$ co- and counter-rotating modes),~\cite{Berti:2003}, and~\cite{Cook:2014cta}. Their low-spin behavior discussed in more detail below is well compatible with the higher-accuracy investigation of~\cite{Cook:2014cta} in particular --- while the low-spin range is too limited in~\cite{Onozawa_1997,Berti:2003} to unambiguously compare the trends.

We provide our complex frequencies and angular separation constants results for these modes as three tables, each one listing a range of $|a_f|$ values with the corresponding $\mathrm{Re}[\omega_{lmn}]$, $\mathrm{Im}[\omega_{lmn}]$, $\mathrm{Re}[\mathscr{A}_{lmn}]$ and $\mathrm{Im}[\mathscr{A}_{lmn}]$ for $M_f$ set to $1$. We make these tables available here:~\cite{my-codeberg, my-github}, along with the \xj{codes} that we used. Given the convergence criterion mentioned above, we consider each of these values to be accurate up to the number of digits provided, that is, to an absolute precision of $10^{-10}$ for each of these quantities. These results complement the data of~\cite{berti:2005ys,berti:2009kk,berti-webpage,vitor-webpage} and they correct and complement the results of the \texttt{qnm} package~\cite{Stein:2019mop} for these three branches. 
We use these tables, along with the \texttt{qnm} package for all other modes, to produce the fits described in Section~\ref{sec:fits}.

The corotating modes in these tables are provided for the ranges $a_f \in [\spinmin{n},1]$ mentioned above, with a step on $a_f$ set to $\delta a_f = 10^{-4}$ and further refined close to $a_f = 1$ ($\delta a_f = 10^{-5}$ for $0.9990 \leq a_f \leq 0.9999$ and $\delta a_f = 10^{-6}$ for $a_f \geq 0.9999$).
For either mode, convergence was extremely slow at and \xj{in the vicinity of} its respective $\spinmin{n}$ spin value, preventing the investigation of a large number of $a_f$ values below this point \xj{or a decrease of the spin step near this point}. This \xj{difficult convergence} is likely a consequence the known failure of Leaver's method in the vicinity of the algebraically special Schwarzschild mode $\omega = - 2 \, \iota$ \cite{chandrasekhar:1975zza,Berti:2003}, where these modes lie at low spins. We could not achieve any convergence --- even at a much lower precision level --- for the few $a_f < \spinmin{n}$ values that we probed (\emph{e.g.}, at \xj{$a_f = \spinmin{n} - \delta a_f$ and $a_f = \spinmin{n} - 2 \, \delta a_f$ in both cases with $\delta a_f = 10^{-4}$, at $a_f = 3 \cdot 10^{-3}$ for $n=8$, or at $a_f = 5 \cdot 10^{-3}$ for $n=9$}). In both cases, $\mathrm{Re} \big[\omega_{lmn}  (a_f = \spinmin{n} ) \big]$ is very close to zero, and extrapolating the frequency solutions to lower spins would make them \xj{cross the imaginary axis at a finite spin value $\spinmin{8} - 2 \, \delta a_f < a_f < \spinmin{8} - \delta a_f$ for $n=8$ and at a finite spin value $\spinmin{9} - \delta a_f < a_f < \spinmin{9}$ for $n=9$, with the imaginary part of the frequency remaining distinct from $-2$ at the crossing point in both cases}. 

While the lack of convergence could simply be due to a complete failure of the method in this range, these results --- including the extrapolated values of $a_f$ and $\mathrm{Im}(\omega_{22n})$ where the imaginary axis would be crossed, although we find them with lower accuracy --- are fully consistent with the findings of~\cite{Cook:2014cta}. With an investigation extended even closer to the imaginary axis,~\cite{Cook:2014cta} indeed found both branches to reach the axis at a finite spin and away from the algebraically special Schwarzschild mode (with $\omega_{22n} \simeq -1.96384 \, \iota$ at $a_f \simeq 3.4826 \cdot 10^{-3}$ and $\omega_{22n} \simeq -2.04223 \, \iota$ at $a_f \simeq 5.3279 \cdot 10^{-3}$ for $n=8,9$, respectively), and to disappear at lower spins. \cite{Cook:2016A} additionally showed that these QNMs do not exist either \emph{on} the imaginary axis itself.

For the counter-rotating mode that we present in a third table (more precisely, this table corresponds to the corotating $n=8$ mode for $m=2$, $a_f < 0$ or \xj{for} $m=-2$, $a_f > 0$, tied to the $m=2$, $a_f > 0$, $n=8$ counter-rotating mode by the symmetry relations~\eqref{eq:flipm}--\eqref{eq:sym}), convergence was also slower at spins very close to $0$ but could still be achieved down to \xj{$|a_f| = 10^{-6}$} --- yet not at the Schwarzschild $a_f = 0$ limit itself, as expected. The values we obtain for $|a_f| > 0$ are however compatible with the Schwarzschild algebraically special limit $\omega_{22n} \rightarrow - 2 \, \iota$ (along with the $(m,n)$--independent Schwarzschild limit $\mathscr{A}_{22n} \rightarrow 4$) for $a_f \rightarrow 0$, to within less than \xj{$5 \cdot 10^{-10}$} by extrapolation. This is in agreement with the analytical prediction of this limit for the counter-rotating mode by~\cite{MaassenvandenBrink:2000}, while the solution \xj{also} obtained down to $|a_f| = 10^{-6}$ in~\cite{Cook:2014cta} was also compatible with it. We accordingly assumed the validity of this limit and added it to the table at $a_f=0$.

However, for this mode, convergence was much poorer at high spin values. We accordingly provide the results for this mode over the spin range $|a_f| \in \xj{ [0, 0.997] }$ (the values at $a_f = 0$ being assumed as mentioned above), with a step \xj{$\delta a_f = 10^{-6}$ at $|a_f| \leq 2 \cdot 10^{-5}$, $\delta a_f = 10^{-5}$ at $2 \cdot 10^{-5} \leq |a_f| \leq 2 \cdot 10^{-4}$}, $\delta a_f = 10^{-4}$ at \xj{$2 \cdot 10^{-4} \leq |a_f| \leq 0.97$, and} $\delta a_f = 10^{-3}$ at \xj{$0.97 \leq |a_f| \leq 0.997$. For both of the next spin values $|a_f| = 0.998$ and $|a_f| = 0.999$, convergence could not be reached, with the algorithm even appearing to be divergent in the second case}.

The disappearance  of the $n= 8,9$ corotating modes --- or at least the failure of the algorithm --- at low spin values prevents the association of the counter-rotating mode to either corotating branch. On the other hand, like the previous studies, we have found only this single counter-rotating solution for any given spin value in this region of the complex plane. Hence, as for the $\omega = - 2 \, \iota$ Schwarzschild QNM, we associate this counter-rotating branch to both corotating ones. Since we label these two corotating branches as the $n=8$ and $n=9$ tones, the associated Schwarzschild mode and counter-rotating branch may be considered as a degenerate $n=8$ and $n=9$ mode simultaneously.

\section{The waveform catalog}
\label{sec:nr_data}
\begin{figure*}[t!]
\subfloat{
\includegraphics[width=0.98\columnwidth]{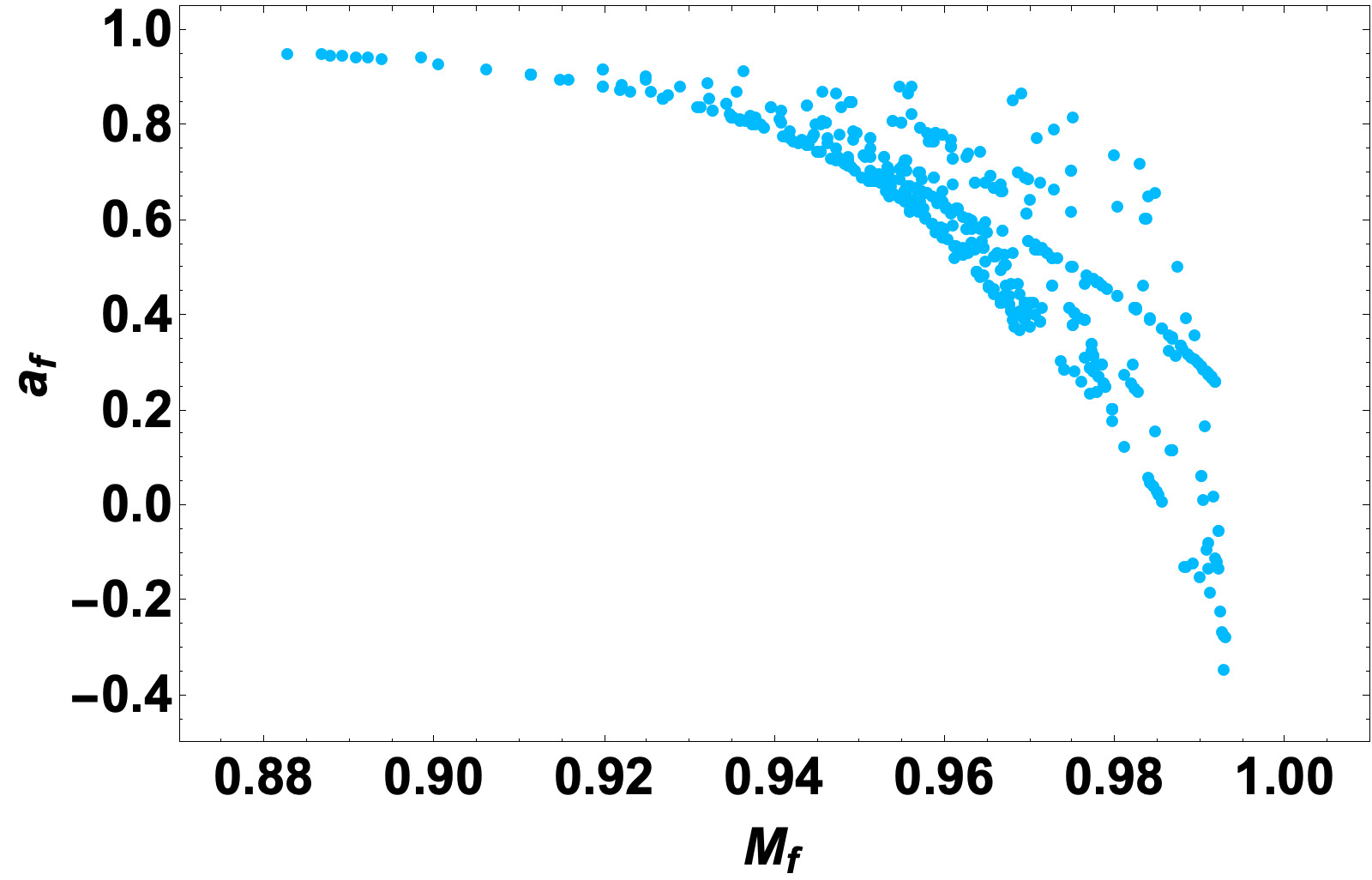}}
\subfloat{
\includegraphics[width=0.98\columnwidth]{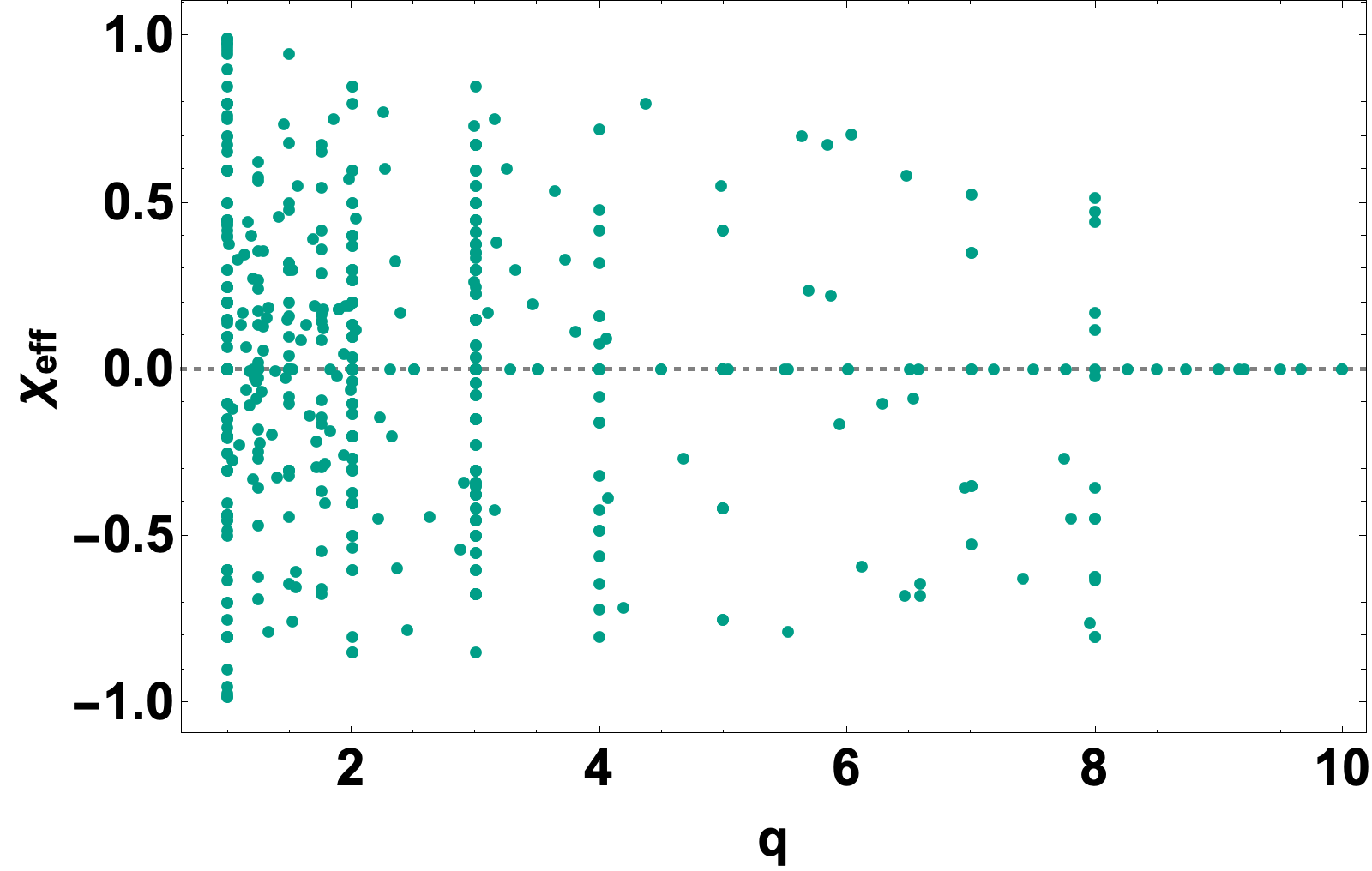}}
    \caption{Left panel: Distribution of the final mass $M_f$ and final spin $a_f$ of the \Nrcases SXS waveforms used in this work. Right panel: same distribution but for the effective spin $\chi_{\mathrm{eff}}$ and the mass ratio $q$ instead.} 
    \label{fig:param_cov}
\end{figure*}
In this work, we have used \Nrcasesall non-precessing waveforms from the SXS catalog~\cite{sxscatalog}, and two from the RIT catalog~\cite{ritcatalog} for comparison. We excluded \Outliers out of the \Nrcasesall SXS cases, which we did not consider accurate enough for our analysis or which had seemingly inconsistent final parameters (see Appendix~\ref{sec:outliers}). We show in Fig.~\ref{fig:param_cov} the parameter space corresponding to the \Nrcases SXS simulations analysed. The left panel shows the $(M_f, a_f)$ distribution, with $M_f \in [ 0.883, 0.993]$ and $a_f\in [-0.344, 0.997]$. Here and in the following, we make the masses dimensionless by setting the total initial mass of the two merging black holes $M = m_1 + m_2$ to unity. On the right panel, we alternatively show the distribution of the same SXS setups in terms of the mass ratio $q=m_1/m_2 \in [1,10]$ and the effective spin $\chi_{\mathrm{eff}}=(\chi_1 m_1 + \chi_2 m_2)/(m_1+m_2) \in \left[-0.97,0.9988\right]$. The visible correlation between $M_f$ and $a_f$ is physical: at fixed mass ratio, the relative energy radiated, $E_f=1-M_f$, increases with the value of $\chi_\mathrm{eff}$ which is itself correlated with $a_f$~\cite{Rezzolla:2007rz,jimenez-forteza:2016oae,hofmann:2016yih}. Among all the cases analysed here, there are \Nrcasesneg cases with $a_f<0$, for which we have used the symmetries given by Eq.~\eqref{eq:sym}.

\subsection{Estimates of the final mass and spin}
The final mass and final spin provided by NR catalogs are usually estimated from their quasilocal definitions~\cite{sxscatalog,szilagyi:2009qz,ashtekar:2004cn,Iozzo2021} on the apparent horizon (AH). In the ringdown regime, the distorted black-hole spacetime evolves quickly towards its stationary state. Then, the black hole spin $S$ is obtained by computing the set of approximate Killing vectors $\phi^i_{(k)}$ and the extrinsic curvature $K_{ij}$ at the AH and integrating them over the induced AH area as,
\begin{equation}
S_{\phi_{(k)}}=\frac{1}{8\pi} \int_{\mathrm{AH}} \phi_{(k)}^i \, s^j K_{ij} \, dA\,,
\end{equation}
where $s^j$ is the outgoing spacelike unit normal vector and $S_{\phi_{(k)}}$ is the spin component estimate along the Killing vector $\phi^i_{(k)}$. The spin magnitude is then evaluated as $S\equiv \sqrt{S_{\phi_{(1)}}^2+S_{\phi_{(2)}}^2+S_{\phi_{(3)}}^2}$. The final mass relies on the spin value $S$ and it is obtained by using the Christodoulou formula for uncharged black holes~\cite{christodoulou:1970wf},
\begin{equation}
    \left( M_f^\mathrm{l} \right)^2 = M^2_{\mathrm{irr}} + \frac{S^2}{4 M_{\mathrm{irr}}^2}\,,
        \label{eq:m_loc}
\end{equation}
which depends on the value of the final spin $S$ and the irreducible mass $M_{\mathrm{irr}}$ (for further details see Sec.2.2 of~\cite{sxscatalog}). We work with the local dimensionless spin, namely,
\begin{equation}
a_{f}^\mathrm{l}= \frac{S}{(M_f^\mathrm{l})^{2}}\,.
\label{eq:a_local}
\end{equation}

The superscripts ${}^\mathrm{l}$ stand for \textit{quasilocal} mass and spin. Alternatively, the mass and the spin can also be estimated from the energy and angular momenta radiated away in the form of gravitational radiation. These \emph{radiation}-based quantities (labeled with a superscript ${}^\mathrm{r}$) are obtained in terms of the Newman-Penrose scalar $\psi_4$,
\begin{align}
\begin{split}
    M_f^{\mathrm{r}}&=\xj{M_{\mathrm{in}}}-\lim_{r\rightarrow\infty} \frac{r^2}{16 \pi}  \int \sum_{lm} \left|\frac{d  h_{lm}}{dt} \right|^2 \, dt'\, ; \\
    a_f^{\mathrm{r}}&=J_\mathrm{in}+\lim_{r\rightarrow\infty} \frac{r^2}{16 \pi}  \, \mathrm{Re}\! \left[\int \sum_{lm} m \, h^*_{lm} \, \frac{d (h_{lm})}{dt}  \, dt'\right]\,,
    \label{eq:m_a_rad}
\end{split}
\end{align}
where $h$ is the gravitational wave strain, $\psi_4=d^2h/dt^2$, and $J_\mathrm{in}$ \xj{and $M_{\mathrm{in}}$ are} the initial ADM dimensionless angular momentum \xj{and initial ADM energy, respectively}. To dissipate the local gauge effects, the radiation quantities are evaluated at a distance $\mathcal{O} (100M)$ away from the black holes apparent horizons (which have a $\mathcal{O} (1M)$ radius) and extrapolated to null infinity. \xj{The integrals are evaluated \xj{starting at a time $t' = t_\mathrm{in}$ which is always taken to be later than the emission of} the junk radiation.}

\subsection{Resolution and extrapolation errors}
\label{sub:NRerror}
Usually, the local grid the near-horizon zone of binary BH simulations is better resolved that the radiation zone so that $a_{f}^{\mathrm{l}}$ and  $M_{f}^{\mathrm{l}}$ are estimated to larger accuracy than $a_{f}^{{\mathrm{r}}}$ and  $M_{f}^{{\mathrm{r}}}$. Moreover, the radiative quantities are also affected by extrapolation errors when extrapolating from $r=\mathcal{O}(100 M)$ to null infinity, by conversion errors from $\psi_4$ to $h$ or by still non-zero residual gauge effects~\cite{jimenez-forteza:2016oae,keitel:2016krm,hinder:2013oqa,Iozzo2021}. Since the fit results presented below in Sec.~\ref{sub:fitvaram} attempt to recover the final mass and spin from radiative quantities (\emph{i.e.}, from the strain mode $h_{22}$) and not from their quasilocal definitions, the errors on these estimates will be better described by the errors on the parameters computed from Eq.~\eqref{eq:m_a_rad} rather than the errors from Eqs.~\eqref{eq:m_loc} and~\eqref{eq:a_local}. Thus, we consider two type of basic errors estimates\footnote{Since most of the simulations used in this work are shown to be in the convergent regime~\cite{sxscatalog}, the main source of errors is either the resolution or the extrapolation to null infinity of the NR datasets. On the other hand, other sources of errors such as the conversion from $\psi_4$ to $h$~\cite{hinder:2011xx} have not been considered here.}: the local error $\delta \epsilon_\mathrm{l}$ as in~\cite{Finch:2021iip} and the radiation error $\delta \epsilon_\mathrm{r}$ that are both defined from the following estimates:
\begin{equation}
\small
   \delta \epsilon_{\mathrm{l},\mathrm{r}} =  \sqrt{\left(\frac{{}^{(N)\!}M_f^{\mathrm{l},\mathrm{r}}-{}^{(N-1)\!}M_f^{\mathrm{l},\mathrm{r}}}{M}\right)^2 +({}^{(N)}a_f^{{\mathrm{l},\mathrm{r}}}-{}^{(N-1)}a_f^{\mathrm{l},\mathrm{r}})^2}\,.
    \label{eq:epsilon_err}
\end{equation}

Here, $M=m_1 +m_2=1$, and the superscripts ${}^{(N)}$ and ${}^{(N-1)}$ stand either for consecutive resolution levels or for consecutive extrapolation orders\footnote{%
One can also use the same formula to compute a mass and spin discrepancy $\delta \epsilon_{\mathrm{l}}$ or $\delta \epsilon_{\mathrm{r}}$ between simulation results from two different codes --- rather than comparing different resolution levels or extrapolation orders within a given code --- as we will do in Sec.~\ref{sub:fitvaram}.
}. Our final radiative error estimates are obtained by combining the resolution and the extrapolation errors, while only resolution effects are relevant for the local estimates (see Sec.~\ref{sub:NRerror}). We have restricted the sum of Eq.~\eqref{eq:m_a_rad} to $l=m=2$, in order to avoid the error contributions sourced by the higher angular modes. Notice that this is only valid to get the error on the fit estimates from the $(22)$ mode but not accurate enough to compute the final mass and final spin. \xj{The same holds for the initial ADM mass and angular momentum in Eq.~\eqref{eq:m_a_rad}: their exact value and time at which they are computed become irrelevant \xj{for the error estimation}, since these terms are suppressed when \xj{computing differences between resolutions or extrapolation orders as per} Eq.~\eqref{eq:epsilon_err}.}
\begin{figure}[]
\includegraphics[width=0.98\columnwidth]{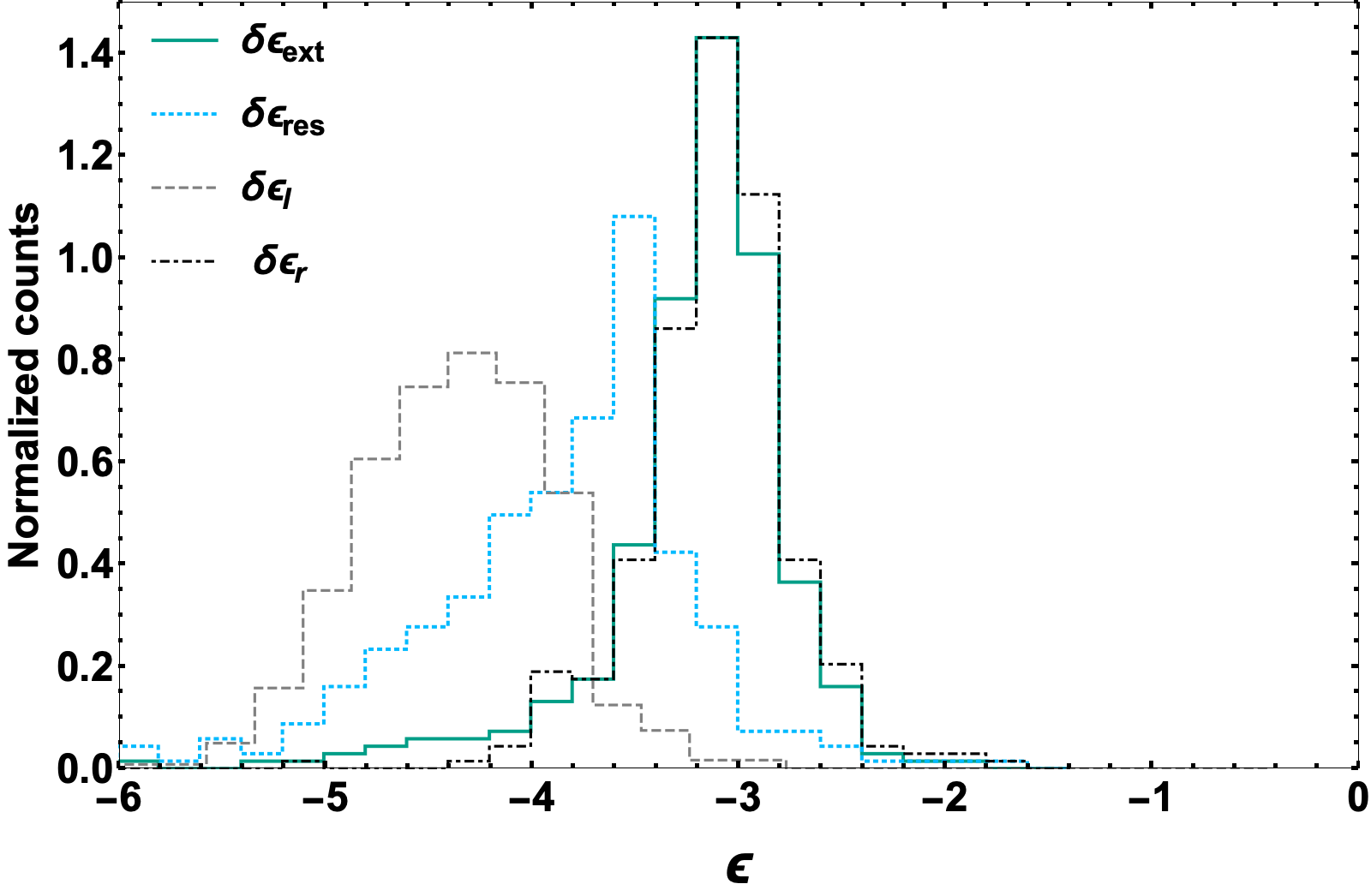}
    \caption{Distributions of error values obtained for the NR local error (dashed gray), the NR radiation error (\xj{dashed-dotted} black) and its extrapolation (\xj{solid} green) and resolution (\xj{dotted} blue) contributions, for the SXS waveforms considered here and when multiple resolutions were available. We observe that the extrapolation errors are typically larger than the resolution ones. The median values obtained are $\left\{1.7\times 10^{-4},7.6\times 10^{-4}, 8.1\times 10^{-4}\right\}$ for \xj{resolution, extrapolation and combined} radiation errors respectively. The distribution of (resolution-based) local errors is additionally shown (\xj{dashed gray} line) and they can be noticed to be substantially smaller than the radiative errors, with a median value $\widetilde{\delta \epsilon_\mathrm{l}} = 4.6\times 10^{-5}$.}
    \label{fig:err_exres}
\end{figure}

More specifically, the local errors have been obtained from the differences on the mass and spin between the highest ($N$) and the second-highest ($N-1$) resolution datum per NR case, following Eq.~\eqref{eq:epsilon_err}. This type of error results from the discreteness of the grid, thus, its value depends on the sampling of the numerical domain. In the case of the local errors, the discreteness affects the computation of the integrals on the AH in Eq.~\eqref{eq:a_local}. Since the resolution is usually finer at the black hole `near-horizon' length scale $\mathcal{O} (1M)$, the local errors are in general smaller than the radiative ones. Moreover, the radiative errors account for all types of inaccuracies that have been propagated to the strain $h_{22}(t)$. Here, we have estimated them from i) the resolution errors $\delta \epsilon_{\mathrm{r},\mathrm{res}}$ and ii) the extrapolation errors $\delta \epsilon_{\mathrm{r},\mathrm{extr}}$. Like for the local errors, the resolution errors are computed from the difference between the highest ($N$) and second-highest ($N-1$) resolutions, but now estimated on the strain\footnote{BH merger simulations are solved by splitting the whole space-time in a set of subdomains that range from the black hole scale to the waves scale, where the black hole scale is usually the finest and the wave scale the coarsest.} $h_{22}(t)$ . The extrapolation errors arise from the extrapolation of the strain to null infinity. Such extrapolation is performed by fitting with second- to fourth-order polynomials the phase and the amplitude of the strain multiplied by radius $r\, h_{22}(t)$ extracted on a set of several finite $\mathcal{O} (100M)$ distances from the black hole local domain~\cite{sxscatalog,keitel:2016krm,jimenez-forteza:2016oae, Iozzo2021}. Here, we estimated the associated error also from Eq.~\eqref{eq:epsilon_err} by taking the differences between the successive extrapolation orders $N=2$ and $N=3$ on the waveform $h_{22}(t)$ at the highest resolution\footnote{A lower-order polynomial typically performs better at extrapolating the ringdown regime (see Sec.~2.4.1 of~\cite{sxscatalog}), hence we did not consider the fourth order, and we always take the second-order extrapolation level as the default $h_{22}(t)$}. The final radiative error is estimated as,
\begin{equation}
    \delta \epsilon_\mathrm{r} =  \sqrt{(\delta \epsilon_{\mathrm{r},\mathrm{res}})^2+\delta \epsilon_{\mathrm{r},\mathrm{extr}}^2} \, .
\end{equation}
In Fig.~\ref{fig:err_exres} we show the distribution of the errors  $\delta\epsilon_\mathrm{l}$ and $\delta\epsilon_\mathrm{r}$ as well as the separate contributions $\delta \epsilon_{\mathrm{r},\mathrm{res}}$, $\delta \epsilon_{\mathrm{r},\mathrm{extr}}$ for the SXS waveforms considered. To compute $\delta\epsilon_\mathrm{l}$, $\delta \epsilon_{\mathrm{r},\mathrm{res}}$ and $\delta \epsilon_\mathrm{r}$, we have only used the \Nrresol simulations that are provided with multiple resolution data. Notice that the extrapolation error (\xj{solid} green) becomes the major contribution to the uncertainty of the SXS dataset used in this work, being typically larger than the resolution one. This can also be seen from the median values of both error estimates: as expected, this value is \xj{larger} for the radiative error, with $\widetilde{\delta \epsilon_\mathrm{r}}=\epserrr$, than for the local error, with $ \tilde{\delta \epsilon_\mathrm{l}} =\epserrl$.

\section{Setup and fit results}
\label{sec:fits}

In this section we show the results obtained from the fits of the $(22)$ mode of the NR waveforms by a range of ringdown models. Each model corresponds to the $(22)$ mode of Eq.~\eqref{eq:rdmodel} where we have simply set $t_r = 0$, and restricted to a total number $\nmax$ of QNM overtones with $\nmax \in \left\{ 0, \dots, 12 \right\}$ --- and occasionally up to $\nmax=16$. In particular, we consider two main scenarios: i) the final mass and final spin are fixed and set equal to the known NR values; or ii) we seek for the mass and the spin that minimise the fit mismatch $\mathcal{M}$ (see below).

In these two scenarios, we have neglected the counter-rotating modes of Eq.~\eqref{eq:rdmodel} since they are expected to have negligible amplitudes compared to the corotating ones~\cite{forteza2020,Finch:2021iip} and to have a negligible impact on the recovery of the final mass and spin~\cite{Dhani:2021vac}. Similarly, mode-mixing effects are as well discarded due to their small impact on the $(22)$ mode~\cite{Finch:2021iip,cook2020,Dhani:2021vac}.

\subsection{The fitting algorithm}
\label{sub:fitalg}
In scenario i), we fit for the $2(\nmax+1)$ parameters $\vec{\lambda}=\lbrace A_{22n}, \varphi_{22n}\rbrace$ with $n \in \left\{ 0, \dots, \nmax \right\}$, for known final mass and spin; while in scenario ii) we fit as well for the mass $M_f$ and the spin $a_f$, while in scenario ii) we fit for the same set of parameters $\vec\lambda$ over a range of $(M_f,a_f)$ values and then optimize the results over this mass and spin range, thus accounting for $2(\nmax+2)$ parameters in total.

Notice that once the values of the mass and spin are fixed, the RD ansatz~\eqref{eq:rdmodel} is linear in the complex amplitudes $\mathcal{A}_{lmn}$. Therefore, one may use a linear least-squares algorithm to obtain the fit results~\cite{giesler2019,Bhagwat:2019dtm,cook2020}.
That is, for a given value of the $(M_f,a_f)$ pair, the complex amplitudes $\mathcal{A}_{lmn}$ are obtained by minimising the $\chi^2$,
\begin{equation}
\label{eq:chi2}
\chi^2=\sum_{k} \left|\bar{h}_{22}\!\left(\vec{\lambda}\right)(t_k)-h_{22}(t_k)\right|^2,
\end{equation}
where the subscript $k$ labels the values of the time axis of the NR waveform, $t_k \in \left[t_0, t_f \right]$ for a certain fit starting time $t_0$ and with $t_f = 90 M$; and $\bar{h}_{22}(\vec{\lambda})$ denotes the model $(2,2)$-mode strain for the set of parameters $\vec\lambda$. By default in the following, the starting time is set to $t_0 = 0$, which corresponds to the peak of the $(22)$ mode of the strain $h_{22}(t)$. We however let this value vary in Secs.~\ref{sub:fitresults} and~\ref{sub:varyt0} as specified there. The best-fit parameters per RD model $\vec{\lambda}^\mathrm{bf}(\nmax)$ are chosen as the ones that minimise Eq.~\eqref{eq:chi2}.

The above fully describes the fitting procedure in scenario i). In scenario ii), the same process is iterated over a range of $(M_f,a_f)$ values to find the optimal one. To this aim, we build a two-dimensional adaptive grid on the final mass $M_f$ and the final spin $a_f$, with a grid minimum step set to $3.2 \cdot 10^{-6}$ in both variables. Every point of the grid is then treated as a linear least-squares minimization problem on the parameters $\vec{\lambda}=\lbrace A_{22n}, \varphi_{22n}\rbrace$ as above~\cite{Finch:2021iip,my-codeberg, my-github}.
Closely related to the $\chi^2$ and recurrently used in GW astronomy, we  compute the mismatch for each best-fit RD model, namely\footnote{In this framework, one can easily show  that both $\mathcal{M}$ and the $\chi^2$ provide the same qualitative behavior. In particular, for a model closely fitting the NR waveform, ${\chi^2 \simeq 2\,\mathcal{M} \, (\sum_k | h_{22}(t_k)|^2)}$. Therefore, a minimum on $\chi^2$ directly translates to a minimum in $\mathcal{M}$ and \emph{vice versa}.},
\begin{equation}
    \mathcal{M} = 1 - \frac{\left\langle h_{22} \, \middle| \, \bar{h}_{22}\left(\vec{\lambda}^\mathrm{bf}\right)\right\rangle}{\sqrt{\Big\langle h_{22} \, \Big| \, h_{22}\Big\rangle \left\langle \bar{h}_{22}\!\left(\vec{\lambda}^\mathrm{bf}\right) \, \middle| \, \bar{h}_{22}\!\left(\vec{\lambda}^\mathrm{bf}\right)\right\rangle}}\,,
    \label{eq:mismatch}
\end{equation}
where 
%%%
\begin{equation}
    \langle f|g\rangle = \int_{t_0}^{t_f} f(t) \, g(t)^* \, dt \, .
\end{equation}
Finally, the best-fit mass and spin values $(M_f = M_f^\mathrm{fit},a_f = a_f^\mathrm{fit})$ for the given waveform and the given number of overtones $\nmax$ of the RD model are selected as the grid point where $\mathcal{M}$ from Eq.~\eqref{eq:mismatch} is minimal.

The associated value of the minimum $\mathcal{M}$ for each RD $\nmax$ model is sufficient to assess the fit accuracy but insufficient to determine whether the fitting parameters are physically reliable. A decreasing value of the mismatch $\mathcal{M}$ between different models is particularly sensitive to overfitting, especially if it is applied to nested models such as the RD models we have considered in this work (the RD model with $\nmax-1$ overtones corresponds to the subclass of the RD model with $\nmax$ overtones with $\mathcal{A}_{\nmax}$ set to $0$). To overcome this issue we use the mass and spin bias $\epsilon$ defined in Eq.~(4) of~\cite{giesler2019},
\begin{equation}
    \label{eq:epsilon}
    \epsilon = \sqrt{\left(\frac{\delta M_f}{M}\right)^2 +\delta a_f^2} \; ,
\end{equation}
where $\delta M_f = M_f^{\mathrm{fit}}-M_f^{\mathrm{true}}$ and $\delta a_f = a_f^{\mathrm{fit}}-a_f^{\mathrm{true}}$. Thus, $\epsilon$  measures the combined deviation of the final mass $M_f$ and the final spin $a_{f}$ with respect to the true parameters $M_f^{\mathrm{true}}$ and $a_f^{\mathrm{true}}$ of the NR simulation, that are estimated from the mass and spin quasilocal definitions~\cite{sxscatalog}, \emph{i.e.}, following Eqs.~\eqref{eq:m_loc} and~\eqref{eq:a_local}.

\subsection{Fits with the mass and spin fixed to their true values}
\label{sub:fitresults}
\begin{figure}[]
\includegraphics[width=0.98\columnwidth]{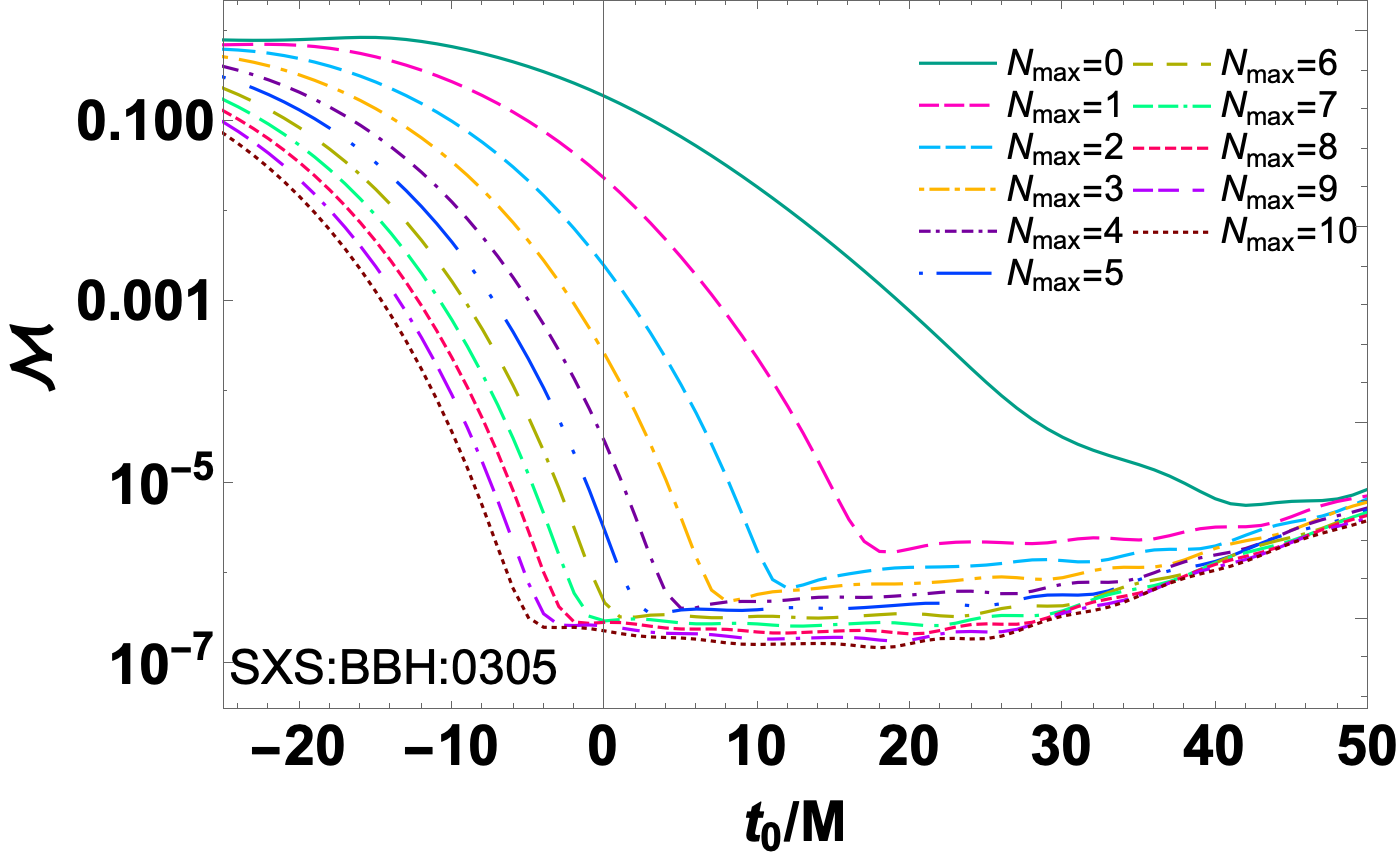}
    \caption{We show the mismatch ${\mathcal{M}}$ at best-fit complex amplitudes $\mathcal{A}_{22n}$ from Eq.~\eqref{eq:mismatch} for a range of RD models with $\nmax \in \left\{ 0, \dots, 10 \right\}$ as a function of the fit starting time $t_0/M$. Notice that as the number of tones of the model increases, the mismatch decreases progressively. In addition, for each $\nmax$, one finds a local minimum on ${\mathcal{M}}$ as a function of $t_0/M$, which occurs at progressively smaller fit starting times $t_0/M$ as $\nmax$ increases.}
    \label{fig:chi2_time}
\end{figure}

First, we show the results obtained by fitting the RD models~\eqref{eq:rdmodel} to the NR waveform SXS:0305 following the same methodology described in~\cite{Bhagwat:2019dtm,forteza2020,Mourier:2020mwa}. With $a_f^{\mathrm{true}}=0.692$ and $M_f^{\mathrm{true}} = 0.952$, this waveform is consistent with the first gravitational wave event GW150914, and it has been recurrently used in several RD studies~\cite{giesler2019,Bhagwat:2019dtm}. In Fig.~\ref{fig:chi2_time} we show the mismatch curves for a set of models with a number of overtones $\nmax$ spanning $\left\{ 0, \dots, 10 \right\}$. In the RD models~\eqref{eq:rdmodel} used for these fits, the final mass and spin are fixed (scenario i), which implies that the whole set of QNM frequencies and damping times are fixed to their GR values. We analyse the fit results as a function of the fitting starting time $t_0/M$. Notice that the mismatch diminishes as the number of overtones $\nmax$ increases for all RD models and for any fit starting time $t_0$. For each RD model, we find a local minimum in $\mathcal{M}$ as $t_0$ varies, after an initial strong decrease and followed by a plateau of nearly-constant $\mathcal{M}$. This minimum\footnote{%
This first local minimum is the global minimum in $\mathcal{M}$ for $\nmax \leq 6$. For larger $\nmax$ values, the global minimum of $\mathcal{M}$ is different and occurs at a later fit starting time $t_0 \simeq 20M$, but it is still only marginally smaller that the first local minimum.
} occurs at increasingly early starting times $t_0$ as the number of overtones $\nmax$ increases. In particular, it occurs at $t_0 \simeq 0$ for the $\nmax=7$ model as it has been observed in~\cite{giesler2019,Bhagwat:2019dtm,Finch:2021iip}. 
For the new $\nmax=8, 9, 10$ models, the local minimum in mismatch occurs at some $t_0<0$. The same trend continues for all the subsequent models tested in this work (that is, with ${10 < \nmax \leq 16}$), which have not been included here for the sake of the plot clarity.

At large $\nmax$, the decrease in the mismatch value (hence also of the $\chi^2$) with increasing $\nmax$ may be mostly due to overfitting of the data. Namely, increasing $\nmax$ increases the number of free parameters in the model accordingly, which induces a decrease in $\mathcal{M}$ and may become the main source of the observed decrease as $\nmax$ gets large. In the next section we discuss how $\epsilon$ may be used as an approximate indicator to detect the overfitting in our RD models. 

\subsection{Varying the final mass and spin}
\label{sub:fitvaram}
 
\subsubsection{Results comparison for SXS/RIT waveform pairs}
\label{subsub:sxsrit}
We now set $t_0=0$ and we study the behavior of $\epsilon$ for two pairs of NR simulations, (\text{SXS:0305}, \text{RIT:0062}) and (\text{SXS:0259}, \text{RIT:0118}). Each pair corresponds to merger simulations from two different catalogs (the SXS catalog~\cite{sxscatalog} and the RIT catalog~\cite{ritcatalog}) with consistent values of every physical parameter. The true values of the final mass and spin are $M_f=\left\lbrace 0.952 \pm  1.2\times 10^{-5},0.966 \pm  3\times 10^{-5}\right\rbrace$ and $a_f=\left\lbrace 0.692 \pm 1.2\times 10^{-4}, 0.581 \pm 2\times 10^{-5} \right\rbrace$ for the first and second pair of simulations respectively, which correspond to merging binary BHs with mass ratio $q=\left\lbrace 1.22, 2.5  \right\rbrace$ and effective spin $\chi_\mathrm{eff}=\left\lbrace -0.0165, 0  \right\rbrace$ respectively. The values of uncertainty quoted on $(M_f, a_f)$ are computed from the differences on the local final mass and spin as $\left|(M_{f}^\mathrm{l})_\mathrm{SXS}-(M_{f}^\mathrm{l})_\mathrm{RIT} \right|$ and $\left|(a_{f}^\mathrm{l})_\mathrm{SXS}-(a_{f}^\mathrm{l})_\mathrm{RIT} \right|$, that translate to a local discrepancy $\delta \epsilon^\mathrm{l}_{\mathrm{SXS}-\mathrm{RIT}} = \left\lbrace 1.2\times 10^{-4}, 3.6\times 10^{-5}  \right\rbrace$ respectively for the two simulation pairs.

\begin{figure}[]
\subfloat{
\includegraphics[width=0.98\columnwidth]{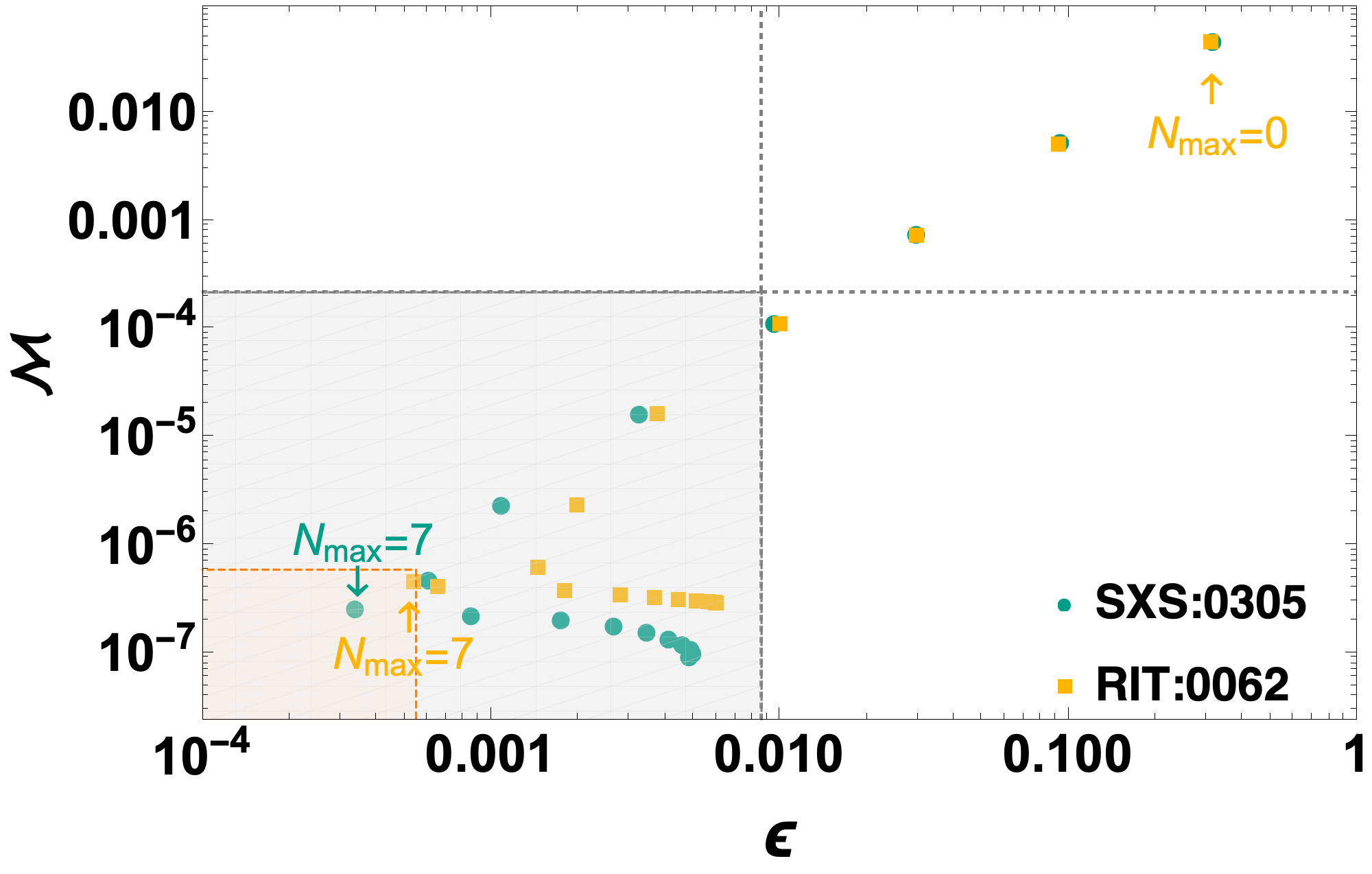}}\\
\subfloat{    
\includegraphics[width=0.98\columnwidth]{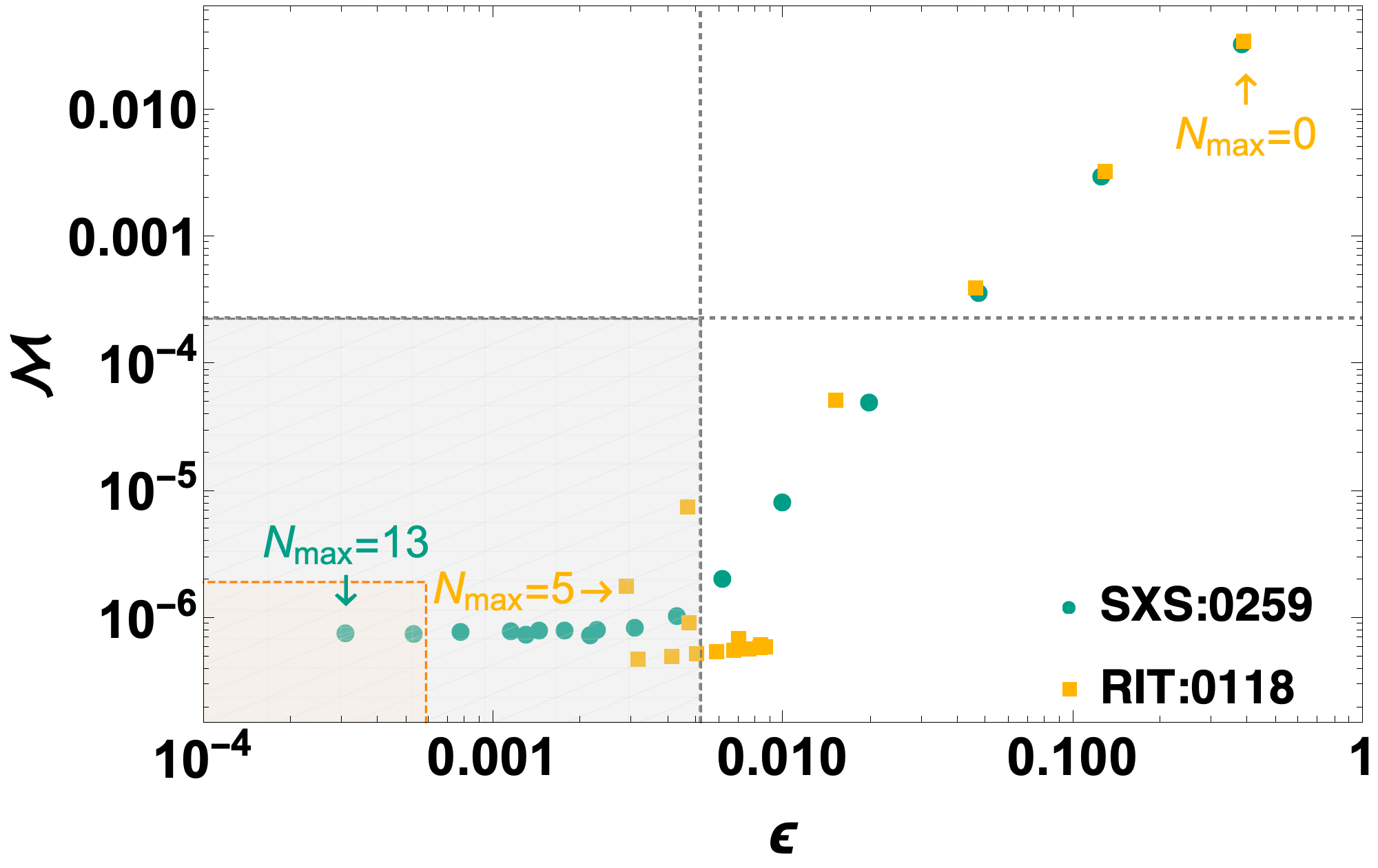}}\\
    \caption{We show the best-fit ${\epsilon-\mathcal{M}}$ plane for a range of models with $\nmax \in \left\{0,\dots,16\right\}$ and for two NR waveforms from the SXS catalog (green) and two waveforms from the RIT catalog (orange), with initial parameters $(q= 1.22,\chi_\mathrm{eff} = -0.0165)$ (top panel) and  $(q= 2.5,\chi_\mathrm{eff} = 0)$ (bottom panel). The gray lines and shaded areas on each panel delimit the mismatch and $\epsilon$ values that are respectively smaller than the mismatch $\mathcal{M}_{\mathrm{SXS}-\mathrm{RIT}}$ and radiative discrepancy on the mass and spin $\delta \epsilon_{\mathrm{r},\mathrm{SXS}-\mathrm{RIT}}$ between the two waveforms considered in the panel. The smaller orange shaded areas at the lower-left corners show the same in terms of the maximum resolution/extrapolation mismatch ${\rm max} \left(\mathcal{M}_\mathrm{res},\mathcal{M}_\mathrm{extr}\right)$, and of the radiative error $\delta \epsilon_\mathrm{r}$, of each SXS waveform. The two green and orange points furthest on the upper-right corner of each panel correspond to the $\nmax=0$ model. In the top panel, notice that as $\nmax$ increases the ${\epsilon-\mathcal{M}}$ points are progressively shifted to the left bottom corner until $\nmax=7$, where the minimum $\epsilon$ is achieved for both NR simulations. Beyond $\nmax=7$,  $\mathcal{M}$ keeps decreasing ---albeit more marginally--- while $\epsilon$ increases. Conversely, for the case shown in the bottom panel,
    we observe that the trend and the values at which $\epsilon_\mathrm{min}$ is achieved are significantly different between the two simulations. In particular, $\epsilon$ now reaches its minimum at $\nmax=13$ for the SXS simulation while the much larger minimal value of $\epsilon$ for the RIT waveform is reached at $\nmax=5$. The difference on the true parameters between the waveforms from both codes is $\delta \epsilon_{\mathrm{l},\mathrm{SXS}-\mathrm{RIT}}\lesssim 10^{-4}$, thus much smaller than the radiative errors $\delta\epsilon_{\mathrm{r},\mathrm{SXS}-\mathrm{RIT}}$ (gray vertical lines).} 
    \label{fig:epsi_chi2}
\end{figure}

In Fig.~\ref{fig:epsi_chi2} we show on a log--log scale, the values obtained for ${\epsilon}$ and ${\mathcal{M}}$ for the two pairs of simulations and for a set of RD models with $\nmax \in \left\{0, \dots, 16\right\}$. On the top panel, corresponding to the first simulation pair (\text{SXS:0305}, \text{RIT:0062}), we observe that both ${\epsilon}$ and $\mathcal{M}$ decrease as $\nmax$ increases up to $\nmax=7$, where $\epsilon$ reaches a minimum at $\epsilon\sim 3\cdot 10^{-4}$ as observed in~\cite{giesler2019,Finch:2021iip}.  This has been considered as one possible empirical evidence that i) one needs $\nmax=7$ overtones to describe the post-peak data~\cite{giesler2019,Finch:2021iip} and ii) post-peak nonlinearities are subdominant even at $t_0=0$. At $\nmax>7$ and for both waveforms, the mismatch keeps decreasing at a reduced rate, while $\epsilon$ increases. This trend reaches a saturation point at $\epsilon\sim 5 \cdot 10^{-3}$ for $\nmax \sim 16$. Thus, in this case, we do not improve any further the accuracy on the estimate of the mass and spin beyond $\nmax=7$, for both SXS:0305 and RIT:0062. This could be taken as the threshold point beyond which overfitting could be significantly affecting the fits. However, this behavior is rather variable when studying other NR cases (see bottom panel and its discussion below, and Sec.~\ref{subsub:epsilondistr}), where one finds that the minimum $\epsilon$ point is case-dependent. On the other hand, in Sec.~\ref{sub:amplitudes} we show that large instabilities in the best-fit amplitudes could be affecting the tones at $n\geq 2$, thus, any claim about the onset point of overfitting shall be taken with caution.

The shaded gray area delimits the domain for which the mismatch is lower than the mismatch between the two waveforms $\mathcal{M}_{\mathrm{SXS}-\mathrm{RIT}}$, and where $ \epsilon\leq \delta \epsilon_{\mathrm{r},\mathrm{SXS}-\mathrm{RIT}}$. Conversely, the smaller shaded orange area near the lower-left corner of each plot stands for the radiation error on the SXS data alone (since only one resolution level per case is provided for the RIT catalog); see Sec.~\ref{sub:NRerror} for further details on the computation of $\delta\epsilon_{\mathrm{r},\mathrm{l}}$. In particular, the upper bound of the SXS error on the mismatch axis --- the mismatch horizontal orange line --- is estimated as the maximum mismatch that results from comparing both the two highest resolution and the two best extrapolation levels ($N=2$ and $3$), namely, ${\rm max} \left(\mathcal{M}_\mathrm{res},\mathcal{M}_\mathrm{extr}\right)$.

In the bottom panel we show the $\mathcal{M}-\epsilon$ results for the higher mass ratio pair (SXS:0259, RIT:0118). Notice that the trend on $\epsilon$ changes substantially compared to the previous case. As expected, the mismatch always decreases, but flattens out at $\nmax=6$, especially for SXS:0259. Here however, for SXS:0259, the value of $\epsilon$ decreases to eventually reach its minimum only at $\nmax=13$ with $\epsilon\sim 3\cdot 10^{-4}$ --- a similar minimal value as the one reached at a smaller $\nmax$ in the case of SXS:0305. For $\nmax>13$, $\epsilon$ increases again although it is not yet saturated at $\nmax=16$. In the case of RIT:0118, $\epsilon$ decreases and
hits its minimum at $\nmax=5$ with a larger value $\epsilon\sim 3\cdot 10^{-3}$ to thereafter grow, saturate at $\epsilon \sim 10^{-2}$ for $\nmax \sim 9$ and decrease again. These discrepancies between the behaviors observed for both waveforms on the lower panel arise in the $\mathcal{M}\leq \mathcal{M}_{\mathrm{SXS}-\mathrm{RIT}}$, $ \epsilon\leq \delta \epsilon_{\mathrm{r},\mathrm{SXS}-\mathrm{RIT}}$ domain, thus they could be affected the by NR errors --- or other systematics --- of each code. 

It is moreover noteworthy that the minimal $\epsilon$ values --- or the turning points on the $\mathcal{M}-\epsilon$ plane --- occur close the boundary delimited by the SXS radiative error $\delta\epsilon_\mathrm{r}$ (orange areas) for both SXS waveforms analysed here. This could indicate that the change of trend for $\nmax \geq \nmax(\epsilon_\mathrm{min})$ for the SXS models may be dominated by the NR uncertainties.

\subsubsection{Mass and spin recovery biases for the set of non-precessing SXS simulations}
\label{subsub:epsilondistr}
\begin{figure}[]
\includegraphics[width=0.98\columnwidth]{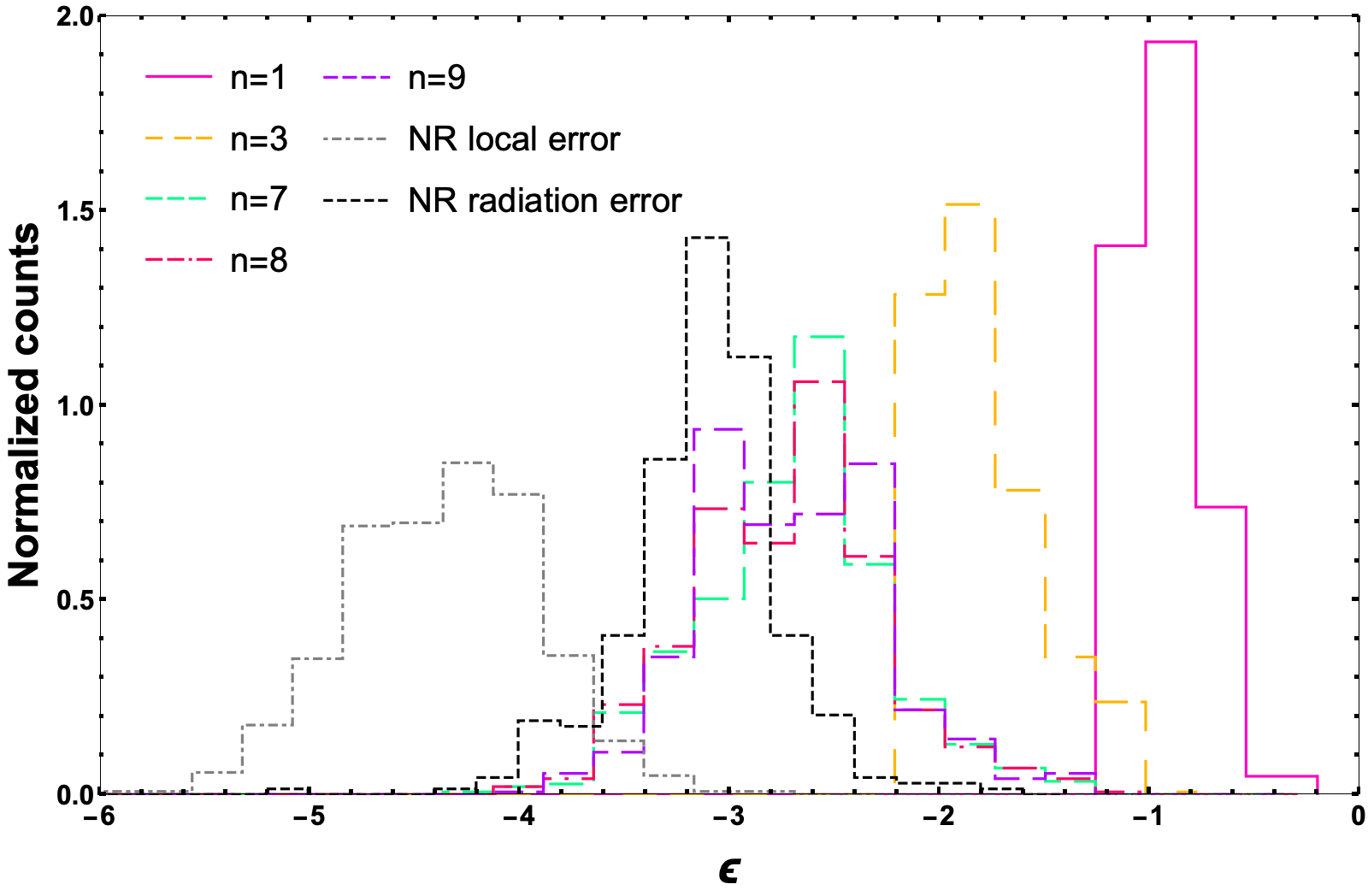}
    \caption{We show the ${\epsilon}$ distribution for five models with $\nmax \in \{1,3,7,8,9\}$ computed from the set of non-precessing NR waveforms from the SXS catalog, together with the distribution of NR error estimates for the waveforms for which multiple resolutions are available. The models with $\nmax=1,3$ show significantly larger values for the $\epsilon$ distributions, with $\tilde\epsilon =10^{-1},1.3\cdot 10^{-2}$ respectively. On the other hand, we have obtained $\tilde{\epsilon}\sim \epshigh$ for the $\nmax=7,8,9$ models, thus not showing significant differences among them.} 
    \label{fig:epsi_all}
\end{figure}

Once studied individually the above two NR cases, we extend this analysis to the set\footnote{%
We excluded \Outliers out of the \Nrcasesall such waveforms in the catalog, which did not appear reliable enough for our analysis. We list these cases in Table~\ref{tab:removed_cases} along with the reasons of their exclusion.
} of non-precessing SXS binary-black hole waveforms~\cite{sxscatalog}. In particular, we want to explore whether the consideration of $\nmax>7$ models allows us to find which number of overtones is statistically preferred over this set of NR waveforms. In Fig.~\ref{fig:epsi_all} we show the distributions of the values obtained for $\epsilon$ over the \Nrcases waveforms considered, for each of five RD models with $\nmax \in \{1,3,7,8,9\}$. Consistently with the particular cases shown on Fig.~\ref{fig:epsi_chi2}, we find that among these models the distributions for $\nmax=1$ and $\nmax=3$ provide the largest values of $\epsilon$, with the median values $\tilde{\epsilon}\simeq\epsone$ and $\tilde\epsilon\simeq\epsthree$ respectively, while the distributions on $\epsilon$ for the $\nmax=7,8,9$ models are shifted to significantly lower values. For instance, we have obtained a median value $\tilde{\epsilon}=\epsseven$ for $\nmax=7$, consistent with~\cite{giesler2019,Finch:2021iip}. We do not observe significant differences between the $\nmax = 7,8,9$ models, where all three distributions overlap within the $10$--$90$ percentiles. We also show on this figure the distribution of the NR radiative ($\delta \epsilon_\mathrm{r}$) and local ($\delta \epsilon_\mathrm{l}$) errors. For the radiative errors, we have taken into account the resolution and extrapolation errors\footnote{274 out of the \Nrcases SXS waveforms discussed here are only available at a single resolution. Thus, these cases have not been accounted for in our NR error estimates.}. Notice that the distribution on the NR local errors does slightly overlap with the $\nmax = 7,8,9$ distributions. On the other hand, the radiative errors broadly overlap with the $\nmax=7,8,9$ distributions of $\epsilon$. As described in Sec.~\ref{sub:NRerror}, we have obtained $ \widetilde{\delta\epsilon_\mathrm{r}}\simeq\epserrr$ for the radiative error, thus a slightly smaller but comparable value to $\tilde{\epsilon}(\nmax=7)$. In contrast, we have obtained a much smaller median value, $\widetilde{\delta \epsilon_\mathrm{l}}\simeq\epserrl$, for the local error. 

\begin{figure}[]
\includegraphics[width=0.98\columnwidth]{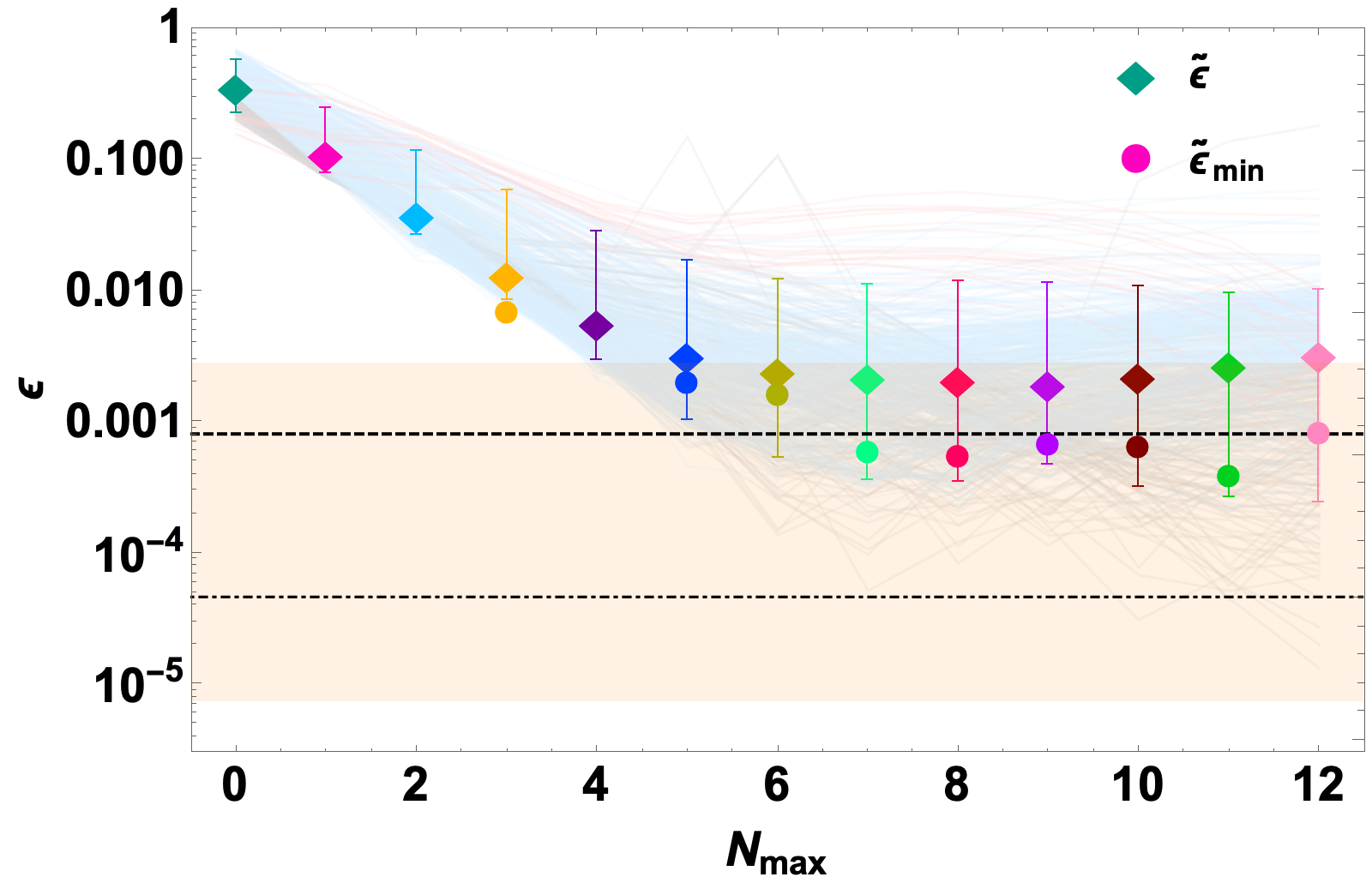}
    \caption{We show the median value ${\tilde{\epsilon}}$ (diamonds) of the distribution of $\epsilon$ values over the SXS waveforms studied, as a function of the number of overtones $\nmax$ included in the model. For each $\nmax \geq 3$, a second value is shown (circles) which has been computed from the median $\tilde \epsilon_{\mathrm{min}}$ of all the studied NR cases for which $\epsilon$ is minimum at $\nmax$ among the $\nmax=0, \dots, 12$ models shown here (there are no such cases for $\nmax < 3$, and a single such case for $\nmax=3$). The error bars represent the $10$--$90$ percentiles of the $\epsilon$ distribution obtained for each of the respective $\nmax$ models. The values obtained for $\epsilon$ as a function of $\nmax$ for each of the \Nrcases cases considered in this work are also shown individually as shaded colored curves. The different colors of these curves represent the cases belonging to different classes of final spin values (see details in the main text, Sec.~\ref{subsub:epsilondistr}). We observe that the median values reach an approximate plateau regime at ${\nmax \sim 5-6}$ and ${\nmax \sim 7}$ for $\tilde \epsilon$ and $\tilde \epsilon_{\mathrm{min}}$ respectively. The shaded orange band shows the $10$--$90$ percentiles of the radiation error distribution, with the dashed black horizontal line denoting the median value $\widetilde{\delta \epsilon_\mathrm{r}}$. The dotted black horizontal line accounts for the median value for the local error, $\widetilde{\delta \epsilon_\mathrm{l}}$.}
    \label{fig:epsi_all_wb}
\end{figure}

In Fig~\ref{fig:epsi_all_wb} we show the $\epsilon$ estimates (shaded colored curves) for the \Nrcasesn~cases we have analysed as a function of the number of overtones $\nmax$ of each RD model. The corresponding median values $\tilde{\epsilon}$ (diamonds) and the $10$--$90$ percentiles of the distribution (bars) are also shown for each $\nmax$. The shaded curves are split in terms of the final spin value as follows: $a_f>0.7$ in light gray,  $0 \leq a_f<0.7$ in light blue and the \Nrcasesneg cases with $a_f<0$ in light red\footnote{%
Since we restrict our models to the corotating modes, we did use in our fits the prior knowledge of the sign of the true parameter $a_f^\mathrm{true}$ to select those corotating modes adapted to this sign. For $a_f^\mathrm{true} < 0$, these modes may then be directly obtained from the positive-spin \emph{counter-rotating} solutions \emph{via} the symmetry relation~\eqref{eq:sym}. Note that in this case $\nmax$ is still to be understood as the total number of distinct overtones in the model, so that the $\nmax = 10$ model for instance will be comprised of the $n=0,1,\dots,7,8,10,11$ $a_f>0$ counter-rotating branches in our notations, since there is no distinct $n=9$ such branch (see Sec.~\ref{sub:qnmresults}). For the consistency of the mode selection, we restrict the allowed spin range on the $(M_f,a_f)$ grid to spins of the same sign as $a_f^\mathrm{true}$ --- either $a_f \in [0,0.99]$ or $a_f \in [-0.99,0]$. In a few cases where $a_f^\mathrm{true}$ is close to $0$ (with either sign) and with small $\nmax$ values (when $\epsilon$ is large), this may lead to an underestimated $\epsilon$ when the minimum-$\mathcal{M}$ solution lies at the $a_f = 0$ boundary of the allowed spin range.
}. First, notice that all $\epsilon$ curves with $a_f \geq 0$ (light gray and light blue) show a similar trend as we vary the number of overtones $\nmax$. This allow us to disregard possible artifacts originating from the mode mixing, i.e., those possible artifacts that result from decomposing the NR strain in terms of spherical rather than spheroidal harmonics, and that would be expected to mostly arise at high spins~\cite{cook2020,Finch:2021iip}. On the other hand, we do observe a higher concentration of the negative spins (light red curves) at high $\epsilon$. \xj{For comparison, at $\nmax = 8$ as an example, the median value of $\epsilon$ over the negative-spin cases (light red curves) is $\tilde{\epsilon}_{a_f < 0} \, \sim 2.0 \cdot 10^{-2}$, substantially larger than the medians of the intermediate positive-spin cases ($0 < a_f < 0.7$, corresponding to the light blue curves) and of the high-spin cases ($a_f > 0.7$, corresponding to the light gray curves), $\tilde{\epsilon}_{0 < a_f < 0.7} \, \sim 2.4 \cdot 10^{-3}$ and $\tilde{\epsilon}_{a_f > 0.7} \, \sim 1.2 \cdot 10^{-3}$ respectively.} We have found that the radiative errors $\delta\epsilon_\mathrm{r}$ obtained for $a_f<0$ are the largest among the NR setup, hence the NR uncertainties could explain the high $\epsilon$ values obtained \xj{in those cases}. \xj{The \xj{minimum-}mismatch value \xj{also} deteriorates for $a_f<0$ with respect to $a_f \geq 0$ cases, for $\nmax \geq 5$. \xj{Beyond this $\nmax$, the median of the best-fit mismatch values over all negative-spin cases approximately plateaus at $\widetilde{\mathcal{M}}_{a_f < 0} \, \sim 4 \cdot 10^{-6}$, while for intermediate positive spins ($0 < a_f < 0.7$) and high spins ($a_f > 0.7$) the corresponding medians only saturate at larger $\nmax$ and reach lower values, $\widetilde{\mathcal{M}}_{0 < a_f < 0.7} \, \sim 2 \cdot 10^{-7}$ and $\widetilde{\mathcal{M}}_{a_f > 0.7} \, \sim 7 \cdot 10^{-8}$ respectively for these two positive-spin classes.} The source of these discrepancies may lie on the numerical setup of the NR simulations and its full exploration may require some further investigation.} On the other hand, we find that $\tilde{\epsilon}$ decreases  before flattening out at ${\nmax \sim 5,6}$. Beyond this point, the values of $\tilde \epsilon$ remain approximately stable at $\tilde \epsilon \sim \epshigh$. The \Nrcasesneg cases with  $a_f<0$ do not increase substantially the value of $\tilde \epsilon$ since they still represent a small fraction of the NR simulations studied. However, this may need to be reviewed if more NR simulations with negative final spin are added to the catalog. The orange shaded area in the lower half of the plot accounts for the $10$--$90$ percentiles obtained from the radiative error distribution while the dashed black line stands for its median value. Note that the $90$th percentile (upper bound) of this error lies above the median values $\tilde \epsilon$ of the $\epsilon$ distribution for $\nmax \geq 6$ (with the exception of $\tilde\epsilon(\nmax=12)$ lying slightly above this line), which suggests that the waveform inaccuracies could be affecting the estimates of $\tilde \epsilon$ at high $\nmax$. Furthermore, it is noticeable that the median $\epsilon$ values for the $\nmax=3-4$ models lie within the $10$--$90$ percentile bands of the $\nmax \geq 5-6$ models. For comparison, the black dotted line shows the median of the local error estimate $\widetilde{\delta {\epsilon}_\mathrm{l}}$. This value is far below the estimates obtained for $\epsilon$, but we recall here that the radiative error $\widetilde{\delta {\epsilon}_\mathrm{r}}$ provides a conceptually more appropriate measure of the error since it is computed directly from the (22) mode of the strain, \emph{i.e.}, from the data used to compute our fits.

In addition, we present on the same figure a second estimate $\tilde{\epsilon}_{\mathrm{min}}$ for each value of $\nmax \geq 3$ (circles), computed as the median of $\epsilon$ over all the cases for which this value of $\nmax$ minimizes $\epsilon$ among the $\nmax \in \{ 0, \dots, 12 \}$ models considered in this analysis. We have not found any case among the SXS waveforms considered for which $\epsilon$ reaches its minimum at an $\nmax < 3$. The value of $\tilde{\epsilon}_{\mathrm{min}}$ for each $\nmax$ is smaller than $\tilde\epsilon$, since all the cases for which $\epsilon$ is not at its minimum at $\nmax$ have been excluded from the distribution in computing $\tilde{\epsilon}_{\mathrm{min}}$. Similarly to $\tilde\epsilon$, the values of $\tilde{\epsilon}_{\mathrm{min}}$ decrease with $\nmax$ before approximately stabilizing for ${\nmax \gtrsim 7}$ at $\tilde{\epsilon}_{\mathrm{min}}\simeq \epshighmin$, which is smaller than the value of $\tilde{\epsilon}$ for those models. Nevertheless, these values of  $\tilde{\epsilon}_{\mathrm{min}}$ lie within the ${\epsilon}$ $10$--$90$ bands and well within the radiative error ${\delta \epsilon_\mathrm{r}}$ distribution.

\begin{figure}[]
\includegraphics[width=0.98\columnwidth]{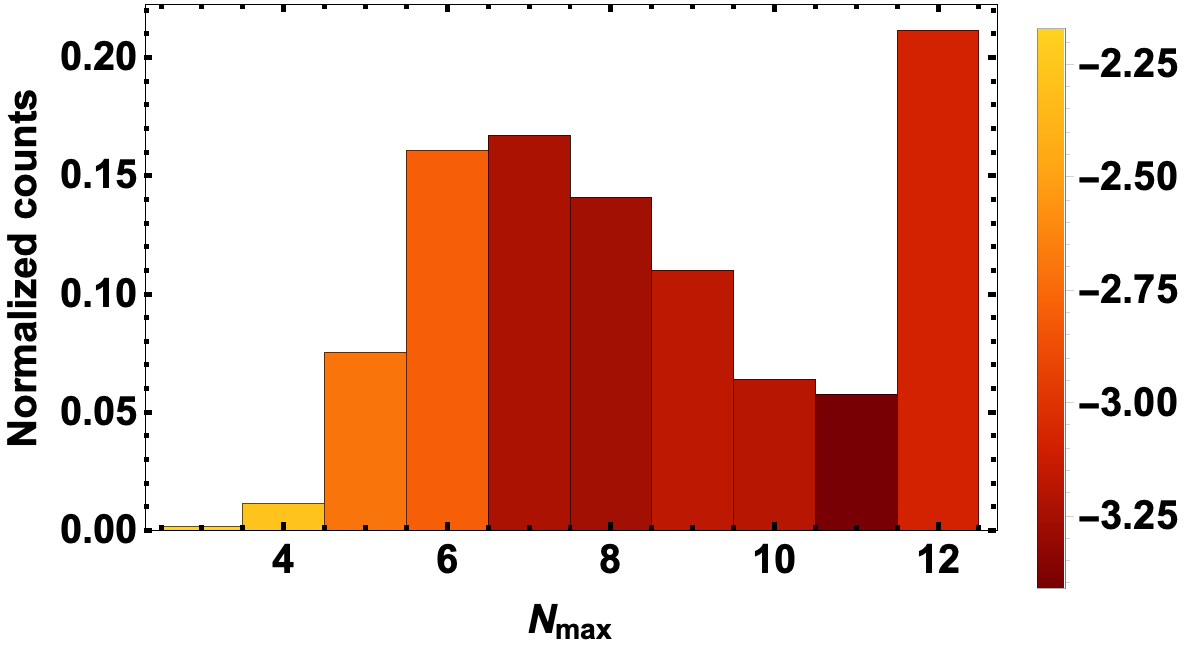}
    \caption{Fraction of the NR cases studied at which $\epsilon$ reaches its minimum value (among the $\nmax \leq 12$ models) at a number of overtones $\nmax$, as a function of $\nmax$. The color function provides the logarithm of the median values over the cases selected in this way $\log_{10}(\tilde\epsilon_\mathrm{min})$ for each $\nmax$ as in Fig.~\ref{fig:epsi_all_wb} (displayed as circles there). We have not found any model reaching its minimum $\epsilon$ at any $\nmax < 3$.}
    \label{fig:histogram}
\end{figure}

Finally, in Fig.~\ref{fig:histogram} we show in histogram form, the fraction of cases for which $\epsilon$ reaches its minimum at $\nmax$ --- among the $\nmax \in \{ 0, \dots 12 \}$ models considered for this figure ---, for each $\nmax$. Here, the color map recalls in log scale the median value of $\epsilon$ among the corresponding cases for each $\nmax$, that is, the $\log_{10}$ of the same values $\tilde\epsilon_\mathrm{min}$ as indicated by circles on Fig.~\ref{fig:epsi_all_wb}. The bulk of the distribution peaks at $\nmax\sim 7$ although we observe that a large fraction (about $21\%$) of the cases have a minimum $\epsilon$ at $\nmax=12$. These include many cases which do not actually reach their minimum $\epsilon$ within the range $\nmax \leq 12$ considered here and that would ideally require to be fit with $\nmax>12$ overtones, such as SXS:0259 reaching its minimum $\epsilon$ at $\nmax=13$ as shown on the lower panel of Fig.~\ref{fig:epsi_chi2}. On the other hand, the corresponding median value of $\epsilon$ for these cases, $\tilde{\epsilon}_{\mathrm{min}}(\nmax=12)$, is moderately higher than the values obtained for cases that reach their minimum earlier, $\tilde{\epsilon}_\mathrm{min}(\nmax=7,\dots,11)$, and is still compatible with the NR error estimates given by $\delta\epsilon_\mathrm{r}$. Therefore, and based on the trend observed for $\tilde\epsilon_\mathrm{min}$ in Fig.~\ref{fig:epsi_all_wb} and on the examples of Fig.~\ref{fig:epsi_chi2}, we do not expect these values to get significantly smaller at $\nmax>12$.

\subsubsection{Stability of the fit amplitudes}
\label{sub:amplitudes}
\begin{figure}[]
\includegraphics[width=0.98\columnwidth]{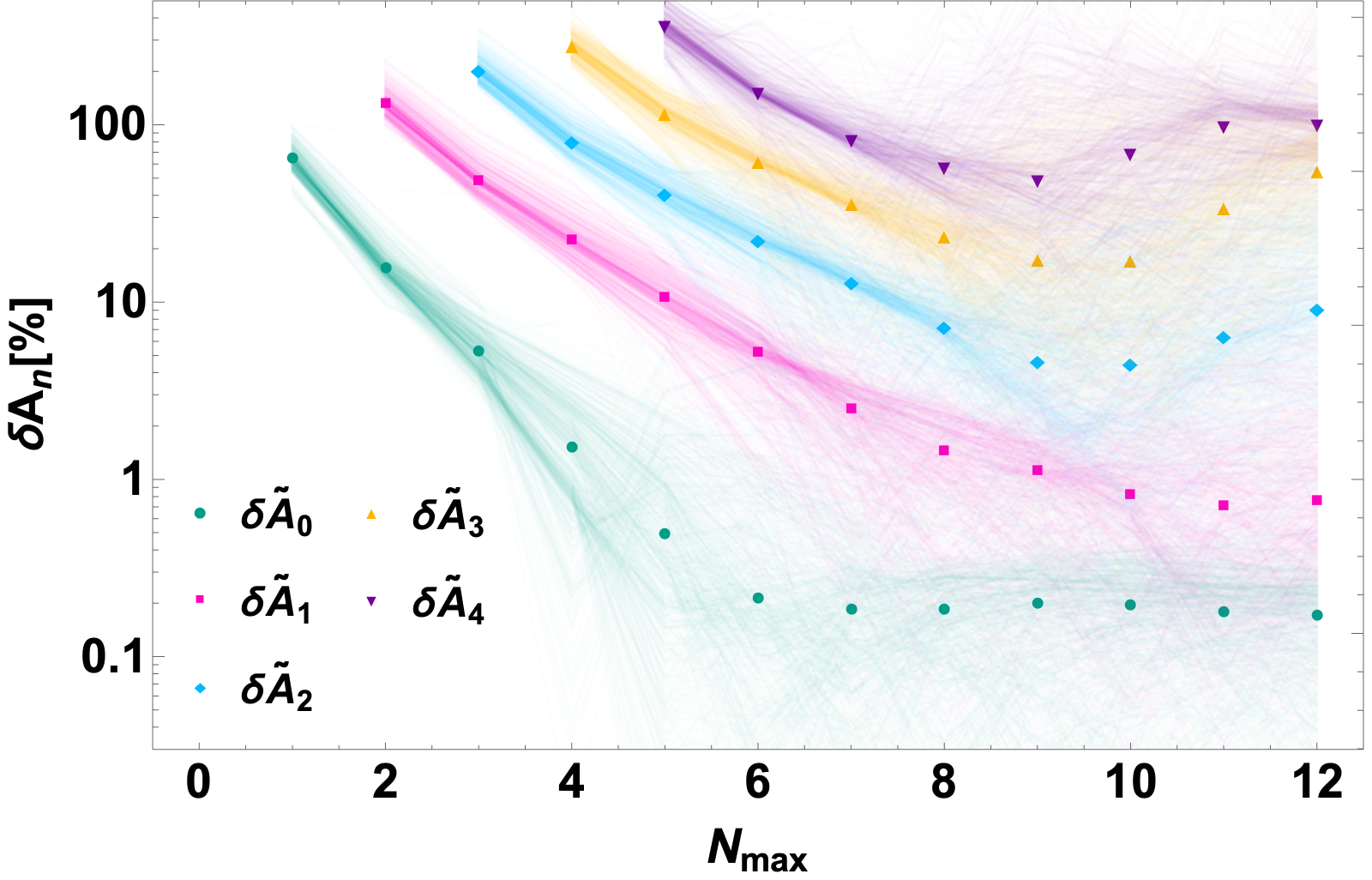}
    \caption{In this figure we show the relative variation of the best-fit amplitudes of each tone $\delta {A}_n(\nmax) = \left| {A}_n^{\nmax}-{A}_n^{\nmax-1} \right|/{A}_n^{\nmax-1}$ as a function of the number of overtones $\nmax$ included in the model, for the $\Nrcases$ SXS waveforms used in this work (shaded colored curves) and for $n=0, \dots, 4$. As we have done for $\tilde{\epsilon}$ in Fig.~\ref{fig:epsi_all_wb}, for each tone $n$, the dots represent the median values $\delta \tilde{A}_n$ per $\nmax$ model.  Each color corresponds to a given tone index $n$, consistently for the median values and the individual curves.  Since we evaluate the differences of the best-fit amplitudes $A_n$ between the consecutive $\nmax$--overtone and $(\nmax-1)$--overtone models, the first point we can evaluate for each $\delta {A}_n$ curve corresponds to the variation with respect to the $\nmax=n+1$ model, $\delta {A}_n(\nmax=n+1)$.}
    \label{fig:amplitudes}
\end{figure}

We study the behavior of the recovered (best-fit) amplitudes of the first five tones $A_n^{\nmax}$, $n=0,...,4$, as we increase the number of overtones $\nmax$ of our models. We require as a criterion for a stable recovery of a given tone $n$, that $A_n^{\nmax}$ remain approximately constant as we modify the number of overtones $\nmax$. In Fig.~\ref{fig:amplitudes}, we show in log scale the relative variation (in percent) of the best-fit amplitudes of each tone $\delta A_n(\nmax) =  \left| A_n^{\nmax}-A_n^{\nmax-1} \right| /A_n^{\nmax-1}$ between successive models as a function of $\nmax$. The shaded colored curves stand for the estimates of $\delta A_n$ for each of the $\Nrcases$ SXS simulations used in this work while the dots stand for the median value $\delta \tilde{A}_n$ of $\delta A_n$ for each ringdown model with a number $\nmax$ of overtones.

For the fundamental mode, we observe that $\delta \tilde{A}_0$ decreases exponentially with $\nmax$ before stabilizing at $\nmax \geq 6$. Remarkably, the flattening of the curve is very similar to the one observed for $\tilde\epsilon$ on Fig.~\ref{fig:epsi_all_wb}. Beyond $\nmax = 6$ the median relative variation of the amplitude $A_0$ remains nearly constant with $\delta \tilde{A}_0~\sim 0.2\%$ \xj{. This is consistent with the results shown in Appendix~\ref{sec:stability_t0}, where we have performed a similar analysis (focusing on $\delta \tilde{A}_0$ and $\delta \tilde{A}_1$) but varying the fit starting time $t_0/M$ at fixed $\nmax$}.

On the other hand, while an initial exponentially decreasing trend with $\nmax$ is also observed for all the overtone modes shown here, the relative variation on the overtone amplitudes is larger than that of the fundamental mode, and increases with $n$. For instance, notice that the $n=1$ amplitude typically varies by about $10\%$ at $\nmax\sim 5$ and that this variation only achieves the $\sim 1\%$ level at $\nmax\gtrsim 8$. The deviations become increasingly larger for $\nmax=2,3,4$, where $\delta \tilde{A}_2\gtrsim 5\%$, $\delta \tilde{A}_3\gtrsim 20\%$ and $\delta \tilde{A}_4\gtrsim 50\%$ for all $\nmax$ values. This symptom of instability seems not to be affecting the estimates of $\tilde \epsilon$ for $\nmax\geq 5$, which surprisingly coincides with the flattening of the $\delta \tilde{A}_0$ curve.

\subsubsection{Analysis at different fit starting times $t_0$}
\label{sub:varyt0}
\begin{figure}[]
\includegraphics[width=0.98\columnwidth]{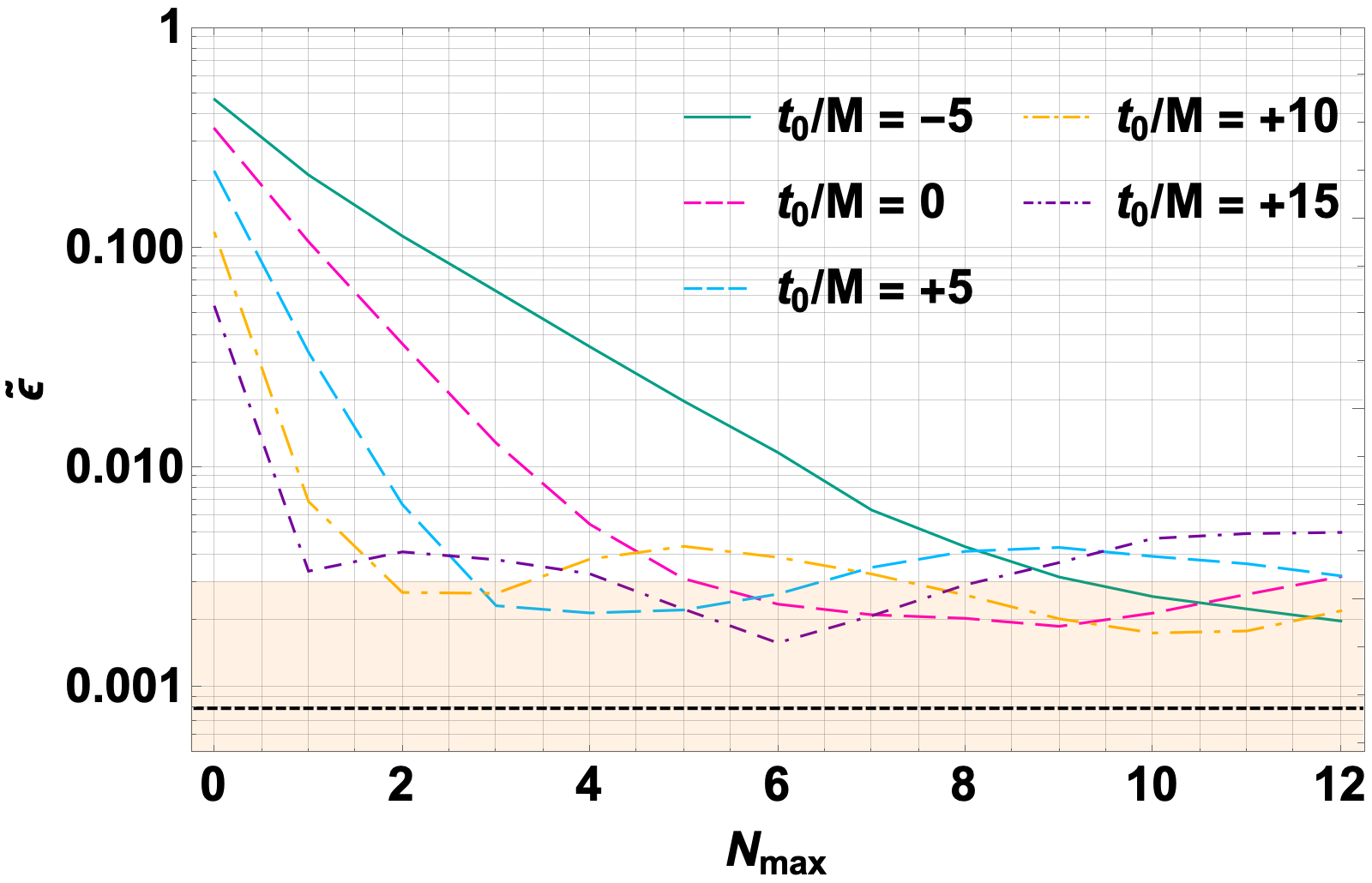}
    \caption{We show the median value $\tilde\epsilon$ of $\epsilon$ as a function of the number of overtones $\nmax$ in the RD model, for a set of normalised fit starting times $t_0/M=\left\lbrace -5,0,5,10, 15\right\rbrace$. Notice that at $t_0/M=\left\lbrace 5,10, 15\right\rbrace$, the curves already nearly hit their minimum at $\nmax \sim \left\lbrace 3,2,1\right\rbrace$ respectively. We show again as a shaded orange band the 10--90 percentiles of the distribution of the radiation error $\delta\epsilon_\mathrm{r}$ (cutting part of the lowest values), as well as its median value as a black dashed line.}
    \label{fig:epsilon_t0s}
\end{figure}

So far, we have studied the NR waveforms using a single starting time $t_0=0$ for all fits as in~\cite{Finch:2021iip,giesler2019}. This specific time corresponds to the time of the peak of the $h_{22}(t)$ strain. However, since this particular time does not have any special physical meaning\footnote{One could reasonably choose $t_0$ as the time at which the final common horizon is formed. However, this time is only well-defined locally while it is causally disconnected from the events happening in the radiation zone. See~\cite{Mourier:2020mwa} and references therein.}, we extend here the analysis for a broader range of fit starting times, with $t_0/M=\left\lbrace -5,0,5,10,15\right\rbrace$. Thus, in Fig.~\ref{fig:epsilon_t0s} we show the median value $\tilde\epsilon$ as a function of the number of overtones $\nmax$ and for the five values of $t_0/M$ selected above.

First, notice that all the curves studied here, show at first a progressively decreasing value for $\tilde\epsilon$ as we incorporate more tones $\nmax$ into the model. This decreasing trend reaches approximately a minimum $\tilde\epsilon \sim (2-3) \cdot 10^{-3}$ (with similar minimum values for all five $t_0/M$ choices shown) at $\nmax \sim \left\lbrace 12, 6, 3, 2, 1\right\rbrace$, for the $t_0/M=\left\lbrace -5,0,5,10,15\right\rbrace$ curves respectively\footnote{%
In the last two cases, $t_0/M = 10,15$, $\tilde\epsilon$ displays two local minima as a function of $\nmax$ within the range considered, with the second minimum being slightly lower than the first. In either case, both minima are nevertheless very similar, so that a value of $\tilde\epsilon$ close to its global minimum is already reached at the first local minimum.
}.

We observe that as the starting time $t_0$ increases, a lower number of tones is required to get close to the minimum value of $\tilde\epsilon$. In particular, the models with $\nmax= 3,2,1$ are appropriate at $t_0/M=5,10,15$ respectively. Furthermore, the fact noted above that the minimum value reached by $\tilde\epsilon$ does not vary significantly with the starting time $t_0$ --- and remains consistent with the distribution of the NR radiative error $\delta\epsilon_\mathrm{r}$ indicated by the orange shaded area --- is particularly intriguing, especially given that this still applies to the curve obtained at $t_0/M=-5$. Notice that at negative times, the amplitude of the strain $(2,2)$ mode is still increasing, so that the morphology of the waveform at such an early stage still differs significantly from a typical exponentially decaying ringdown wave (see for instance Figs.~16-22  of ~\cite{bhagwat:2017tkm}). On the other hand, the fact that $\tilde\epsilon(\nmax=2)_{t_0/M=10}\simeq \tilde\epsilon(\nmax=12)_{t_0/M=-5}$ for instance, suggests that the lower tones $n\lesssim 2$ of the $(\nmax=12)$ model may contribute significantly to the estimates of $\epsilon$ obtained for $t_0/M = -5$ --- with the addition of higher tones into the model helping to better constrain the lower ones.

We justify this behavior relying on two main hypotheses: i) high overtones could be fitting some fraction of the NR noise or ii) high-overtone amplitudes and phases are flexible enough to fit well the early part of the waveform. Hypothesis i) could only be better tested when more accurate waveforms are added to the catalogs. On the other hand, hypothesis ii) would imply that the high-tone degrees of freedom can capture the morphology of the waveform at early times, while the values of $\epsilon$ result predominantly from modelling increasingly better the lower tones $n=0,1..$. This could be the case even irrespective of the NR errors. For instance, the short and similar damping times of the high overtones could induce strong correlations between the higher tones amplitudes combined to a very short time range where they are still measurable, thus impeding one from getting physically reliable information from them. The latter point is also reinforced by the results of the amplitude stability analysis of Sec.~\ref{sub:amplitudes}.

This consideration is also relevant for deciding which starting time and model should be preferred to estimate the final mass and the final spin of a given GW event from its RD phase. As a rule of thumb, the statistical uncertainty on a given parameter $\sigma_\lambda$ scales as the inverse of the RD signal-to-noise ratio (SNR) $\rho\propto 1/\sigma_\lambda$. For an event consistent with GW150914, about half of the total ringdown SNR is lost at $t_0/M=10$ and about $70\%$ at $t_0/M=15$~\cite{bhagwat:2017tkm,Bhagwat:2019dtm,forteza2020}. Therefore, given the minor variation of the minimum $\tilde\epsilon$ with different values of $t_0$ observed here, we expect that for real GW events, only the RD waves with a large number of overtones $\nmax$, at early starting times and that are compatible with or slightly dominated by the statistical error --- i.e., with $\sigma_{\lambda=M_f,a_f} \gtrsim \epsilon(t_0/M,\nmax)$ --- will be appropriate to place as accurate constraints on the mass and the spin as possible.
\section{On the interpretation of the fit models}

The magnitude $\epsilon$ measures systematic deviations on the recovered final mass and final spin with respect to the true parameters of each NR simulation. These deviations may be produced either by i) an insufficient number of tones in our ringdown model, Eq.~\eqref{eq:rdmodel}; ii) numerical errors propagated to the strain $h_{22}(t)$; or iii) the possible presence of nonlinearities in the waveform~\cite{Bhagwat:2019dtm,okounkova:2020vwu}. We have observed in Figs.~\ref{fig:epsi_all} and~\ref{fig:epsi_all_wb} that missing tones on Eq.~\eqref{eq:rdmodel} become the dominant source of deviations on $(M_f, a_f)$ for models with $\nmax \lesssim 4-5$, and that such deviations stabilize on average at $\nmax \gtrsim 6$. Furthermore, there exists some nontrivial correlations between the values of $\tilde \epsilon$ and the amplitudes of the tones. In particular, the fundamental-mode best-fit amplitude $A_0$ stabilises at roughly the same $\nmax$ as $\tilde\epsilon$ does. The amplitudes and phases of the tones are not predicted from the solutions of black hole ringdown perturbation theory but rather fixed by the initial conditions of each NR simulation, \emph{i.e.}, they do not hold any \emph{a priori} unique relation with the final mass and final spin~\cite{berti:2005ys,london:2014cma,forteza2020}. Therefore, the improvement on the $(M_f,a_f)$ estimate at high $\nmax$ is achieved by adding further information to the RD models through the complex frequencies $\omega_{lmn}$, which depend uniquely on the mass and the spin. Concerning the physical reliability of the amplitudes, we observe that the $n>1$ overtones typically suffer from larger than $5\%$ variations on their values when varying the number of overtones $\nmax$ of the model, in agreement with~\cite{forteza2020}. 

As discussed in the main text, with $t_0=0$, the biases on $(M_f,a_f)$ decrease exponentially up to $\nmax \sim {5}$. The gain in accuracy in the estimates of these parameters occurs irrespective of the stability issues observed for the high-overtone amplitudes. Therefore, if one assumes GR to be correct, and assuming a high enough SNR, those models shall provide accurate estimates of the final mass and spin of real GW events, at least in a majority of cases. In line with that argument, one could perform IMR consistency tests, where any inconsistency observed on the final mass and spin (as measured by $\epsilon$) between the inspiral and merger-ringdown regimes could be hinting for new physics as it would suggest a violation of the no-hair theorem. The results obtained in~\cite{giesler2019,Finch:2021iip} and complemented in this work, suggest that an QNM model with ${\nmax \sim 3-6}$ overtones would be able to constrain the final parameters up to $\epsilon\sim 10^{-2}- 10^{-3}$ in many cases (although some signals would only reach such constraints with more overtones). This level of accuracy is beyond the current LIGO-Virgo typical SNR-limited uncertainties on the mass and the spin~\cite{LIGOScientific:2020tif}, which together result in $\epsilon \sim 0.1$. However, this accuracy  may be achieved and surpassed with the third-generation detectors LISA, Einstein Telescope and Cosmic Explorer.

On the other hand, the stability issues observed on the amplitudes, together with the variable $t_0$ analysis, become relevant in order to assess to which extent these models and, in particular, high--overtone number models can be used for performing black hole spectroscopy. This would imply estimating \emph{independently} the frequencies and damping times of \emph{each tone} together with phases and amplitudes. It is likely  that the instabilities observed in the tone amplitudes may become even larger when adding the frequencies and damping times as extra free parameters. This would thus eventually induce systematic errors that may have an important impact on the final estimate of the QNM spectrum itself~\cite{bhagwat:2016ntk,Capano:2021etf,rdownpaper2}.  
%Thus, on average, our newly computed tones $n=8,9$ do help to extend the trends on $\epsilon$ first studied in~\cite{Finch:2021iip,giesler2019} to models with $\nmax>7$. Based on the estimates on $\tilde\epsilon$, we have found that there is no significant improving on the estimates on $M_f$ and $a_f$ for models with ${\nmax\geq 5,6}$ although we can not statistically discard $\nmax=3$ if the 10--90 percentiles are considered. The trend on the $\epsilon$ is slightly  reduced at ${\nmax=8,9}$ if one considers $\tilde{\epsilon}_\mathrm{min}$ though those values are still consistent within the $10$--$90$ $\tilde{\epsilon}$ and error $\delta\tilde{\epsilon}_\mathrm{r}$ percentiles. In summary, the results obtained separately in~\cite{giesler2019,Finch:2021iip} and in this work, suggests that an overtone model with ${\nmax \sim 3-5}$ would be able to constrain up to $\epsilon\sim 10^{-2}- 10^{-3}$. This level of accuracy improves the current LIGO-Virgo estimates on the mass and the spin~\cite{LIGOScientific:2020tif} which would currently translate to $\epsilon \sim 0.4$ although it may not be sufficient for 3G detectors as LISA or ET.  

\section{Conclusions}
The aim of this work has been to study the behavior of high overtones for a set of $\Nrcases$ nonprecessing NR waveforms, extending the fit results obtained in~\cite{giesler2019,Finch:2021iip} to higher than $\nmax=7$ overtones. To this end, we have computed the quasinormal mode frequencies for the overtones with indices $n=8$ and $n=9$ both for the corotating and the counter-rotating branches.

The $n=8$, $n=9$ corotating modes have been computed for a range of spins $a_f\in \left[\spinmin{n},1\right]$ with \xj{$\spinmin{8} = 3.6 \cdot 10^{-3}$ and $\spinmin{9} = 5.4 \cdot 10^{-3}$}, while the one counter-rotating mode associated to these tones is provided for spins ranging from \xj{$a_f = 10^{-6}$} (extended by assumption to $a_f = 0$ with the Schwarzschild $\omega_{228} = - 2 \, \iota$ solution) to \xj{$0.997$}. 
Our results are consistent with~\cite{Onozawa_1997,Berti:2003,Cook:2014cta,Cook:2016A,Cook:2016B} for these three branches, and are made available here:~\cite{my-codeberg}, completing the results provided by~\cite{berti:2005ys,berti:2009kk,berti-webpage,vitor-webpage,Cook:2014cta,cook-QNM,Stein:2019mop}.

First, we have used these results to extend the RD fits to the SXS waveform SXS:0305 to $\nmax >7$, as a function of the fit starting time $t_0$, with the final mass and spin fixed to the simulation's true final parameters. We observe that the mismatch $\mathcal{M}$ keeps marginally decreasing with $\nmax$ for the models with $\nmax=8,9$ and beyond. On this respect, we have found that the first local minimum in mismatch $\mathcal{M}$ as a function of $t_0$ occurs at negative starting times $t_0<0$ for $\nmax \geq 8$, which is possibly due to data overfitting.

Second, we have estimated the value of the final mass -- final spin recovery bias $\epsilon$ for QNM models with $\nmax \in \left\{0,\dots,16\right\}$ overtones, starting at the peak of the $(2,2)$ strain component, $t_0=0$, and for two pairs of SXS and RIT waveforms with identical parameters, $(a_f,M_f)=(0.692, 0.952)$ for the first pair \{SXS:0305, RIT:0062\} and $(a_f,M_f)=(0.581, 0.966)$ for the second pair \{SXS:0259, RIT:0118\}. We have found that the trend on $\epsilon$ can be significantly different between the two simulations and that it is, in general, case-dependent. We estimate a very similar minimum $\epsilon$ of $\epsilon_\mathrm{min} \sim 3\times 10^{-4}$ for the two SXS waveforms, at $\nmax=7$ for SXS:0305 and at $\nmax = 13$ for SXS:0259. For the RIT waveforms with the same parameters, we have obtained  $\epsilon_\mathrm{min} \sim 4\times 10^{-4}$ at $\nmax=7$ and $\epsilon_\mathrm{min} \sim 3\times 10^{-3}$ at $\nmax=6$, for RIT:0062 and RIT:0118 respectively. 

Then, we have applied the fitting algorithm described in Sec.~\ref{sub:fitalg} to \Nrcases out of the \Nrcasesall SXS non-precessing binary black hole simulations and for $\nmax \in \left\{ 0, \dots, 12\right\}$, still with $t_0=0$. Specifically, our results for $\epsilon$ are consistent with the $\nmax=3,7$ models shown in~\cite{Finch:2021iip}. We observe that the median value $\tilde \epsilon$ of the distribution of $\epsilon$ over these simulations decreases exponentially with $\nmax$ to about $\tilde \epsilon \simeq \epshigh$ at $\nmax\sim 5-6$ --- although there is a significant overlap between $\nmax=3,4,5,6$ in the distribution of $\epsilon$ values between the multiple SXS cases. Moreover, $\tilde \epsilon$ does not change significantly beyond $\nmax\sim 6$, which also applies to our new $\nmax=8$, $\nmax=9$ and $\nmax>9$ models. We noted nevertheless that for about $21\%$ of the cases, models with $\nmax \geq 12$ were required to hit the (similar) minimum value on $\epsilon$. The value of $\tilde\epsilon$ appears to always be bounded by the NR errors. We provide optimistic and pessimistic estimates of the NR errors which we have here referred to respectively as i) local errors $\delta\epsilon_\mathrm{l}$ and ii) radiative errors $\delta\epsilon_\mathrm{r}$. The latter should be a more accurate representation of the NR on the strain since it is directly derived from it. We notice in particular that our $90$th percentiles of the radiation error distribution are above the plateau value $\tilde \epsilon \sim \epshigh$, hence they could be affecting or even dominating the $\tilde \epsilon$ values at $\nmax \geq 5$.

Furthermore, we have studied the stability of the best-fit amplitude values $A_n$ for a range of tones $n \in \left\lbrace 0,...,4\right\rbrace$. For the fundamental $n=0$ mode and for the first overtone $n=1$, we have found that the median relative amplitudes variations $\delta\tilde A_0$ and $\delta\tilde A_1$ between successive $\nmax$-overtone and $(\nmax-1)$-overtone models are below the $1\%$ level for RD models with $\nmax \geq 4$ and $\nmax \geq 8$, respectively. On the other hand, we observe that the amplitudes of the overtones with $n>2$ are unstable. We have also found a significant correlation between the typical variation $\delta\tilde A_0$ of $A_0$ as a function of $\nmax$, and the value of $\tilde \epsilon$. These elements could indicate that, for a majority of the studied cases, the improvement on $\epsilon$ is predominantly achieved by increasingly improving the constraints on $A_0$ --- and possibly on the first few overtones to a lesser extent.

Finally, we have repeated the $\nmax \in \{0,\dots,12$\} RD models study over the \Nrcases SXS cases considered over a few different values of the fit starting time $t_0$, with $t_0/M \in \{-5,0,5,10,15 \}$. We have found that the minimum value reached by $\tilde\epsilon$ as a function of $\nmax$ does not vary significantly if we vary the fit starting time $t_0$; a model with $\nmax=7$ overtones at $t_0/M=0$ provides, on average, a similar (slightly higher) accuracy on $\epsilon$ as a model with $\nmax=1$ at $t_0/M=15$. This is relevant since the effects of the overtones $n\geq 1$ are expected to be small at $t_0/M\gtrsim 10$ due to their short damping times. This further supports the hypothesis that the constraints on $\epsilon$  may be predominantly induced by an improvement in modelling the low tones $n=0,1, \dots$, regardless of the weak --- or unstable --- constraints one obtains for the higher tone amplitudes. In this regard, the ans\"atze and, in particular, the higher tones, appear to be sufficiently flexible to accurately fit the strain at times around the peak even when those high overtones take amplitude values that are likely unstable, or even nonphysical.

We note that our ringdown models~\cite{my-github} may be suitable for performing IMR consistency tests for current and next-generation GW observatories, where the gain in sensitivity may allow us to hit and surpass the accuracy levels observed here for numerical data. On the other hand, based on the results of this paper, we are more skeptical about using overtone models of a given $(l,m)$ mode to robustly perform black hole spectroscopy due to the instabilities observed on the amplitudes of the overtones with $n>1$ --- which are likely to propagate to frequencies and damping times when those are added as free parameters. For the first overtone ($n=1$), such amplitude instabilities are reduced  to the $1\%$ level for models with $\nmax\gtrsim 8$. Thus, a two-tone $n=0,1$ spectroscopy may remain possible provided that one considers a large number of additional tones in the model, at the expense of adding a large number of free parameters.
%%%%%%%%%%%%%%%%%%%%%%%%%%%%%%%%%%%%%%%%%%%%%%%%%%%%%%%%%%%%%%%%%%%%%
%%%%%%%%%%%%%%%%%%%%%%%%%%%%%%%%%%%%%%%%%%%%%%%%%%%%%%%%%%%%%%%%%%%%%
\begin{acknowledgements}
We acknowledge the Max Planck Gesellschaft for support and we are grateful to the Atlas cluster computing team at AEI Hannover for their help. The authors are also thankful to Swetha Bhagwat, Collin Capano, Sumit Kumar, Alex Nitz and Paolo Pani for useful discussions and comments on \xj{this paper. We also thank the anonymous referee for insightful comments and suggestions that led to further improvements to this document.}
\end{acknowledgements}
%%%%%%%%%%%%%%%%%%%%%%%%%%%%%%%%%%%%%%%%%%%%%%%%%%%%%%%%%%%%%%%%%%%%%%
\bibliography{biblio.bib}
%%%%%%%%%%%%%%%%%%%%%%%%%%%%%%%%%%%%%%%%%%%%%%%%%%%%%%%%%%%%%%%%%%%%%
%%%%%%%%%%%%%%%%%%%%%%%%%%%%%%%%%%%%%%%%%%%%%%%%%%%%%%%%%%%%%%%%%%%%%
\appendix
\section{Outliers}
\label{sec:outliers}
\begin{table}[h!]
    \centering
    \begin{tabular}{|c|c|}
    \hline
        \ Index \ & Issue  \\
        \hline \\[-2.5ex]
        0002 & \ Large extrapolation error \  \\
        0084 & Large extrapolation error  \\
        0090 & Large extrapolation error  \\
        0091 & Large extrapolation error  \\
        0158 & Large extrapolation error  \\
        0170 & Reported $M_f > 1$  \\
        0171 & Reported $M_f > 1$  \\
        0218 & Large extrapolation error  \\
        1110 & Large extrapolation error  \\
        1134 & Reported $M_f > 1$  \\
        \hline
    \end{tabular}
    \caption{List of the \Outliers out of \Nrcasesall non-precessing SXS binary-black hole waveforms (labeled under the form \sxs{\emph{index}}) that we do not consider in our analysis. These waveforms are excluded either due to an unphysical value being reported for $M_f$ ($M_f > 1$) in their respective metadata files, or due to a large extrapolation error (as measured by a mismatch value $\mathcal{M} \geq 10^{-3}$ between the waveforms provided at extrapolation orders $N=2$ and $N=3$), as per indicated in the second column.}
    \label{tab:removed_cases}
\end{table}
\begin{figure*}[th!]
\subfloat{
\includegraphics[width=0.98\columnwidth]{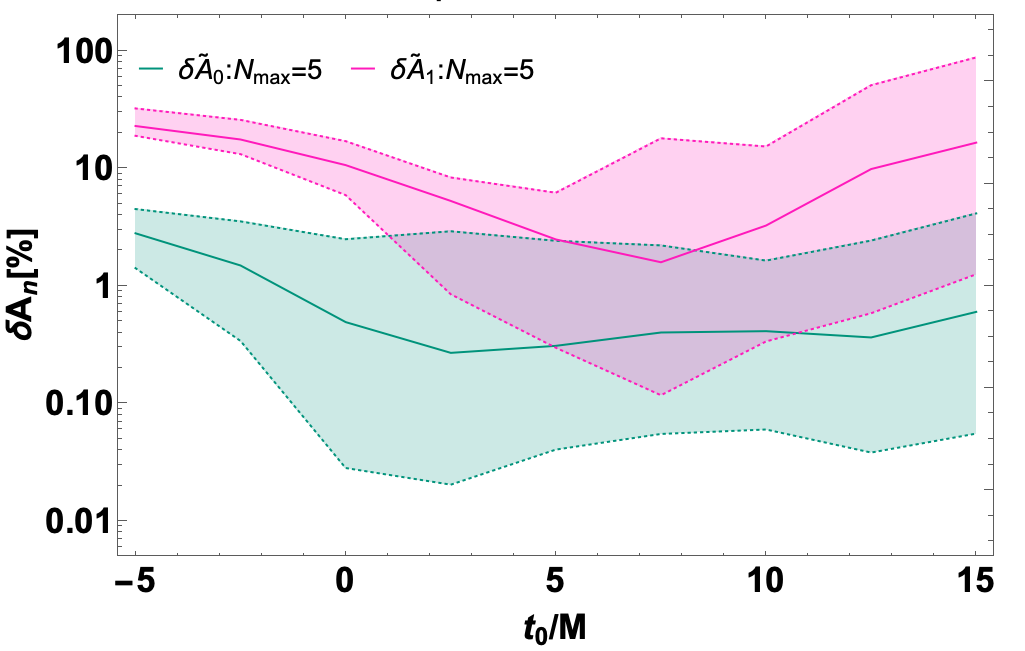}}
\subfloat{
\includegraphics[width=0.98\columnwidth]{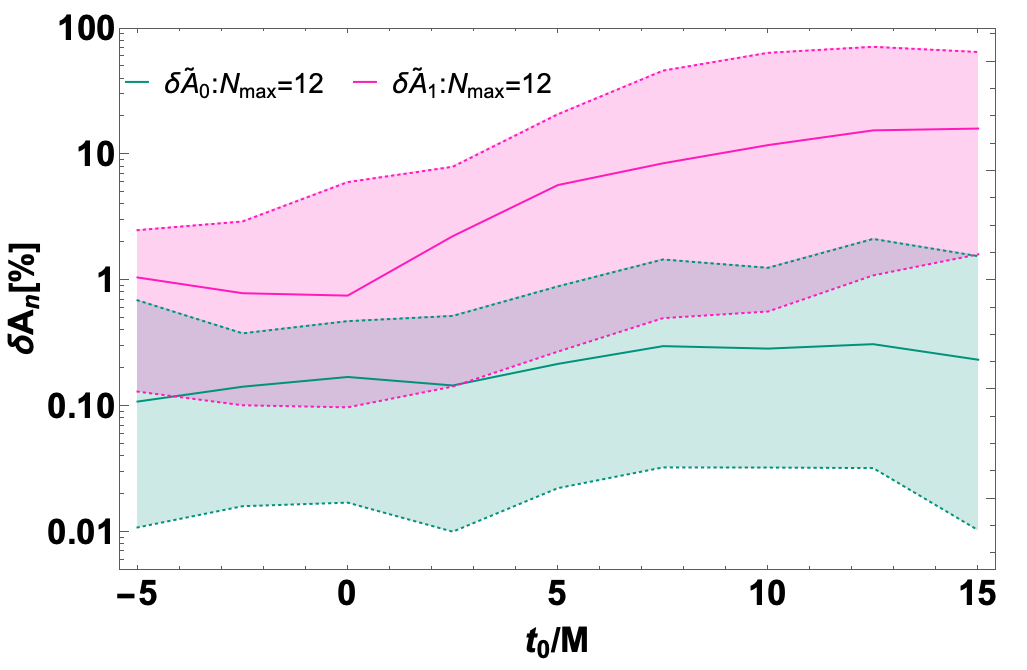}}
    \caption{\xj{Relative variation of the best-fit amplitudes $\delta \tilde{ A}_0(\nmax)$ and $ \delta \tilde{ A}_1(\nmax)$ at $\nmax = 5$ (left panel) and at $\nmax = 12$ (right panel) as a function of the starting time $t_0/M$. The solid curves represent the median values of the \Nrcases cases studied in this work while the shadowed areas stand for the 10--90 percentiles of the corresponding distributions.}} 
    \label{fig:amps_stab_t0}
\end{figure*}
We list in Table~\ref{tab:removed_cases} the \Outliers non-precessing SXS waveforms that we have removed from our analysis. We used the mismatch as a landmark to identify the cases with larger uncertainty. In Fig.~\ref{fig:err_exres} we have seen that the dominant contribution to the numerical uncertainty comes from the extrapolation of the waveform to null infinity. Accordingly, for each of the non-precessing SXS cases, we have computed the mismatch $\mathcal{M}$ as given by Eq.~\eqref{eq:mismatch} between the waveforms with successive extrapolation orders $N=2$ and $N=3$. To this end, we aligned beforehand the two numerical waveforms $h_{22}^{N=2,3}(t)$ in time and in phase so that the peak of the strain is located at time $t=0$ for both, with the same initial phase. We then excluded the cases for which we found an extrapolation mismatch $\mathcal {M}\geq 10^{-3}$. We moreover excluded another three cases with a a seemingly incorrect reported final mass $M_f>1$ (with in fact even $M_f > 2$ in each of these cases).
\section{Stability of the fit amplitudes at varying starting times}
\label{sec:stability_t0}
\xj{In this section we study the respective fractional variations $\delta A_{0}$, $\delta A_{1}$ of the best-fit amplitudes of the fundamental mode and of the first overtone between successive models with $\nmax$ and $\nmax-1$ overtones --- as in Fig.~\ref{fig:amplitudes} --- at different starting times $t_0/M$ and for two separate values of $\nmax$. The first value chosen is $\nmax=5$, \emph{i.e.}, a value at which $\delta \tilde{A}_{0}$ has not yet plateaued at $t_0/M=0$. The second value is $\nmax=12$ for which we expect the values of $\delta \tilde{A}_{0}$ to have plateaued for all the starting times considered in this work, and  $\delta \tilde{A}_{1}$ to have also reached a plateau at $t_0 / M  = 0$}.
\xj{In Fig.~\ref{fig:amps_stab_t0} we show the results for the fractional variations (in percent), with $\nmax=5$ on the left panel and  $\nmax=12$ on the right panel. The solid curves represent the median values $\delta \tilde{A}_{n}$ as shown in Fig.~\ref{fig:amplitudes} while the shadowed areas stand for the the 10--90 percentiles of the distributions over the \Nrcases SXS cases considered in this work. The resolution on $t_0/M$ in this figure is $\Delta (t_0 / M) = 2.5$.}

\xj{For $\nmax=5$, we obtain that $\delta \tilde{ A}_0$ decreases with $t_0$ until $t_0/M \simeq 2.5$, beyond which it approximately stabilizes at $\delta \tilde{A}_0 \sim 0.3\%$ (slightly increasing again towards later times). The larger values at earlier times $t_0/M < 2.5$ are due to the need for a larger number of tones to accurately fit the data --- while the insufficient number of tones induces the further instabilities observed on $\delta \tilde{ A}_0$ (\emph{cf.} Fig.~\ref{fig:amplitudes} at $t_0/M = 0$). On the other hand, the values for $\delta \tilde{ A}_1(\nmax=5)$ decay until reaching a minimum of about $2 \%$ at $t_0/M\sim 7.5$ to thereafter grow back at late times. As for the case of the $n=0$ amplitude, the early decreasing of $\delta \tilde{ A}_1$ is sourced by the lack of higher overtones in the RD model, while the late increase may be due to the effects of the NR noise together with the suppression of the first overtone amplitude at late times (see the further discussion of this tone below).}

\xj{For $\nmax=12$, the values of $\delta \tilde{ A}_0$ remain approximately constant (or very slightly increasing), around $0.1\%$ to $0.2 \%$, for all the starting times studied, where the little variations on their values are consistent with the values of the dispersion of the data given by the 10--90 percentiles. We recover approximately the same values as in the large--$\nmax$ plateau regime as a function of $\nmax$ shown for $t_0/M = 0$ in Fig.~\ref{fig:amplitudes}. Like for $\nmax=5$, we observe an increasing trend for $\delta \tilde{ A}_1(\nmax=12)$, at $t_0/M > 0$. Since here the number of tones is much larger, this indicates that the $n=1$ tone is less stable compared to $n=0$ at late times regardless of the number of tones included in our models. This lack of stability at late times may be originated by the suppression of the first overtone at these times, that is sourced by its high damping factor $\exp(-t/\tau_1)$. Recall that for instance, for a GW150914--like event, $\tau_1/M \sim 3.8$ and the first overtone amplitude is reduced by $\sim {e^{-4}}$ between $t / M = 0$ and $t/M \sim 15$, thus leaving the amplitude values of this tone more exposed to the effects of the NR noise.}
\end{document}